\documentstyle[twoside,aps,prl,multicol,epsf]{revtex}
\begin{document}
%\input REVTeXDraft1.tex
%%%%%%%  MY DEFINITIONS   %%%%%%%%%
\def\B.#1{{\bbox{#1}}}
\def\C.#1{{\cal{#1}}}
\def\BC.#1{{\bbox{\cal{#1}}}}
\def\Ref.#1 {{(\ref{#1})}}
%%%%%%%  GREEK simbols %%%%%%%%%%
\def\a {\alpha}
\def\b {\beta}
\def \bv {\bar\varepsilon}
\def\e {\epsilon}
\def\g {\gamma}
\def\G {\Gamma}
\def\d {\delta}
\def\D {\Delta}
\def\o {\omega}
\def\S  {\Sigma}
\def\s  {\sigma}
\def\k {\kappa}
\def\l {\lambda}
\def\L {\Lambda}
\def\z {\zeta}
\def\t {\tau}
\def\p {\psi}
\def\r {\rho}
\def\oz {\omega_0}
\def\op {\omega_+}
\def\om {\omega_-}
\def\Oz {\Omega_0}
\def\Op {\Omega_+}
\def\Om {\Omega_-}
\def\On {\Omega_n}
\def\O {\Omega}
\def\l {\lambda}
\def\L {\Lambda}
%%%%%%%%%%  EQUATIONS %%%%%%%%
 \def\BE {\begin{equation}}
\def\EE {\end{equation}}
\def\BEA{\begin{eqnarray}}
\def\EEA {\end{eqnarray}}
\def\nn {\nonumber}
\def\nl {\nonumber \\ }    %nn=> newline
%%%%%%%% SOMETHING ELSE %%%%%%%%
\def\la {\langle}
\def\ra {\rangle}
%\renewcommand{\thesection}{\arabic{section}}

%%%%%%%%%%%%%%%%%%%%%%%%%%%%%%

\title{
Anomalous Scaling from Controlled Closure in a Shell Model of Turbulence} 

\author {Victor S. L'vov, Daniela Pierotti, Anna
  Pomyalov and Itamar Procaccia} 

\address{Department of~~Chemical Physics, The Weizmann Institute of
  Science, Rehovot 76100, Israel} \maketitle

\begin{abstract}
  We present a model of hydrodynamic turbulence for which the program
  of computing the scaling exponents from first principles can be
  developed in a controlled fashion. The model consists of $N$
  suitably coupled copies of the ``Sabra" shell model of turbulence.
  The couplings are chosen to include two components: random and
  deterministic, with a relative importance that is characterized by a
  parameter called $\e$. It is demonstrated, using numerical
  simulations of up to 25 copies and 28 shells that in the $N\to
  \infty$ limit but for $0<\e\le 1$ this model exhibits correlation
  functions whose scaling exponents are anomalous. The theoretical
  calculation of the scaling exponents follows verbatim the closure
  procedure suggested recently for the Navier-Stokes problem, with the
  additional advantage that in the $N\to \infty$ limit the parameter
  $\e$ can be used to regularize the closure procedure. The main
  result of this paper is a {\em finite} and closed set of
  scale-invariant equations for the 2nd and 3rd order statistical
  objects of the theory. This set of equations takes into account
  terms up to order $\epsilon^4$ and neglects terms of order
  $\epsilon^6$. Preliminary analysis of this set of equations
  indicates a K41 normal scaling at $\e = 0 $, with a birth of
  anomalous exponents at larger values of $\e$, in agreement with the
  numerical simulations.

\end{abstract}
\vskip .8cm~ \hfill\parbox{9.6cm}{ \small \it It is almost impossible
  to write a paper in the area of fluid turbulence without drawing
  heavily on one or more papers of Bob Kraichnan, who usually
  introduced the issue and wrote one or more fundamental papers
  towards its resolution. The present contribution is not an
  exception.  It uses freely the model that Bob has introduced in
  \cite{70Kra}. Bob, we are happy to dedicate this paper to you at the
  occasion of your 70th birthday. The interaction, intellectual
  stimulation and friendship that you gave us over the last decade are
  unforgettable, and $\,$we $\,$hope $\,$and $\,$trust $\,$that $\,$they
  $\,$will $\,$continue $\,$for $\,$years $\,$to $\,$come. } \vskip .8cm
\begin{multicols}{2}
\section{Introduction}
\label{s:intro} % \vskip -1cm \rightline{\sf s:intro}\vskip   1 cm 
%%%%%%%%%%%%%%%%%%%%%%%%%%%%%%%%
In this paper we present a calculation scheme aimed at evaluating
the scaling exponents that characterize correlation functions of
turbulent fields. For the sake of clarity we consider shell models of
turbulence rather than analyze the Navier-Stokes equations. However we
stress from the start that in principle all the steps provided here
can be repeated in the context of Navier-Stokes turbulence.

Shell models as well as Navier-Stokes turbulence pose an infinite
hierarchy of dynamical equations for the n-order correlation
functions. This hierarchy is {\em linear} in the correlation
functions, and in the limit of infinite Reynolds number is also
homogeneous. It was recently discovered\cite{98LP,98BLP} that this
hierarchy obeys a rescaling symmetry which stems from the rescaling
symmetry of the Euler equation\cite{Fri}. This rescaling symmetry
foliates the space of solutions into slices of different scaling
exponents $h$ of the velocity; these are referred to as $h$-slices.
On each $h$-slice one finds ``normal scaling" with the given value
of $h$. The full solution is a linear combination
of all the solutions on the $h$-slices with non-universal
weights which are determined by the forcing on the integral scale of
turbulence.  Different orders of the correlation functions are
dominated by different $h$-slices, and accordingly the full solution
has anomalous scaling. The anomalous exponents are expected to be
universal.

In trying to evaluate the scaling exponents appearing in this way from
first principles, we proposed\cite{98LP,98BLP,multi-sca} to truncate
the hierarchy of equations, preserving the fundamental rescaling
symmetry that gives rise to anomalous scaling. Truncation is
problematic; in turbulence there is no natural small parameter, and
therefore any closure of an infinite hierarchy is uncontrolled. It is
therefore worthwhile to introduce a 1-parameter family of models,
characterized by a parameter $\e\in [0,1]$, which shows normal scaling
when $\e=0$ and recovers the anomalous scaling of the original model
when $\e=1$.

Following an idea in \cite{70Kra}, we will construct such a family, and show 
that the transition from
normal to anomalous behavior occurs at some finite value of $\e>0$. We
will use $\e$ as a small parameter to regularize the closure
procedure; we will show that our closed equations are valid to
$O(\e^4)$, whereas the neglected terms are of $O(\e^6)$. We can
improve the closure scheme systematically by including terms of
$O(\e^6)$, neglecting terms of $O(\e^8)$, etc.

A way to achieve this small parameter is to couple $N$ copies of the
same turbulent system, be it a shell model or the Navier-Stokes
equations, and choose the coupling to have both a deterministic and a
random contributions, with relative amplitudes $\e$ and
$\sqrt{1-\e^2}$. For $\e=1$ we loose the coupling between the copies,
and recover the initial anomalous problem for any value of $N$. For
$\e=0$ we will show that that in the limit $N\to \infty$ we get normal
scaling. Thus for some value of $\e>0$ and for large enough $N$ we
expect to see the birth of anomalous scaling, hopefully in a
perturbative fashion. The existence of this transition is the main
discovery of this paper, and we study it analytically using the
$\e$-controlled closure procedure, and by direct simulations of the
$N$-copied model.

In Section~2 we review briefly the ``Sabra" shell model of
turbulence\cite{98LPPPV}, and introduce the copy space with the
appropriate coupling. We study the resulting model numerically in
Sect.~3. In Section~4 we start the construction of the theory. First
we present the statistical objects and derive the hierarchy of
evolution equations that they satisfy, exposing in particular their
scale invariance. In Section~5 we discuss the closure which preserves
the scale-invariance, demonstrating how the smallness of $\e$ is used
to control consecutive closure steps.  Sections~6 and 7 are devoted to
analysis of the resulting $\e$-controlled closure, and the results are
compared in Sect.~8 with the numerical findings of Sect.~3. The
paper is on the whole rather technical, and for the sake of the
casual reader we summarize in some detail the points of principle
in the conclusion section which is numbered 9.
%%%%%%%%%%%%%%%%%%%%%%%%%%%%%%%%%
\section{The ($N,\e$)-Sabra model with random couplings}
\subsection{The Sabra model}
\label{ss:sabra}%\vskip -.9cm \rightline{\sf ss:sabra }\vskip   .9 cm 
The starting point is the Sabra shell model as introduced in
\cite{98LPPPV}:
\begin{eqnarray}
{du_n(t)\over
dt}&=&i\Big[ak_{n+1}~u^*_{n+1}u_{n+2}+bk_{n}~u^*
_{n-1}u_{n+1}\nonumber\\
&&-c k_{n-1}~u_{n-2}u_{n-1}\Big]-\nu k_n^2~u_n+f_n(t) \ . 
\label{sabra}
\end{eqnarray}
Here $u_n$ refers to the amplitude associated with ``wave-vector"
$k_n$, where the spacing in this reduced model is determined by
$k_n\equiv k_0 \l^n$; $\l$ is the spacing parameter, $\nu$ the
``viscosity", $f_n(t)$ a random Gaussian force which is operating on
the lowest shells.  The parameters $a,~b$ and $c$ are restricted by
the requirement
\begin{equation}
a+b+c=0 \ , \label{sum}
\end{equation}
which guarantees the conservation of the``energy"
\begin{equation}
E=\sum_{n=0}^N |u_n(t)|^2 \ , \label{energy}
\end{equation}
in the inviscid, unforced limit.

The equations of the Sabra model are invariant under the phase
transformation
\begin{equation}
u_n \to u_n \exp(i\theta_n) \ , \label{phase}
\end{equation}
where the phases $\theta_n$ are restricted by the set of equations
\begin{equation}
\theta_{n-1}+\theta_n=\theta_{n+1} \ . \label{theta}
\end{equation}
Choosing $\theta_1$ and $\theta_2$ arbitrarily, $\theta_n$ is
determined for all $n>2$. Evidently, the physical results of the model
must be independent of the choice of the phases $\theta_1$ and
$\theta_2$. In particular, the only non-zero correlation functions are
those which are independent of the phases $\theta_n$. The non-vanishing
2nd and 3rd order correlators are
\begin{eqnarray}\label{S2}
S_2(k_n)&=&\langle|u_n|^2\rangle,\\ \label{S3}
S_3(k_n)&=&{\rm Im}\langle u_{n-1}u_n u^*_{n+1} \rangle  \ .
\end{eqnarray}
Note that Eq.~(\ref{sabra}) respects additional phase symmetry
\begin{equation}
u_n\to -u^*_n \ . \label{phasesym}
\end{equation}
One of the consequences of this symmetry is that
\begin{equation}
{\rm Re}\langle u_{n-1}u_n u^*_{n+1} \rangle=0 \ ,
\end{equation}
explaining why we only need to consider the imaginary part in
Eq.~(\ref{S3}).  The symmetries of this model were selected
explicitly in\cite{98LPPPV} to give rise to a small number of nonzero
correlation functions, with the aim of simplifying the calculations
presented in the later sections of this paper.
%%%%%%%%%%%%%%%%%%%%%%%%%%%%%%%%%%
\subsection{($N,\e$)-generalization of the Sabra model}
\label{ss:Nesabra}%\vskip -.5cm \rightline{\sf ss:Nesabra }\vskip   .5 cm 
The standard available procedures to generalize dynamical systems to
$N$ copies involve {\em real} variables\cite{61Kra}.  In shell models
in general, and in the Sabra model in particular, the amplitudes $u_n$
are complex. Therefore we rewrite Eq.~(\ref{sabra}), following
\cite{97Pie}, in terms of the real and imaginary parts
$u'_n\equiv$ Re$\{u_n\}$ and $u''_n\equiv$ Im$\{u_n\}$. Doing this we
guarantee, after the generalization to $N$ copies, that the $N\to 1$
limit coincides with the original model. The equations are:
\begin{eqnarray}
&& {du'_n(t)\over
dt}=\Big[\gamma_{a,n+1}(-u'_{n+1}u''_{n+2}+u''_{n+1}u'_{n+2})\nl &&\quad
+ \gamma_{b,n}(-u'_{n-1}u''_{n+1}
+u''_{n-1}u'_{n+1})
+\gamma_{c,n-1}(u'_{n-2}u''_{n-1}\nl &&\quad   +u''_{n-2}u'_{n-1})\Big]-\nu
k_n^2~u'_n+f'_n(t) \,,        \label{sabrareal}\\
&&{du''_n(t)\over
dt}=\Big[\gamma_{a,n+1}(u'_{n+1}u'_{n+2}+u''_{n+1}u''_{n+2})\nl 
&&\quad + \gamma_{b,n}(u'_{n-1}u'_{n+1}+u''_{n-1}u''_{n+1})
 + \gamma_{c,n-1}(-u'_{n-2}u'_{n-1}
\nl &&\quad
+u''_{n-2}u''_{n-1})\Big]-\nu k_n^2~u''_n+f''_n(t) \ , \label{sabraima}
\end{eqnarray}
where 
\begin{equation}\label{defgammas}
\gamma_{a,n}\equiv ak_{n}\,,\quad \gamma_{b,n}\equiv b k_{n}\,,
\quad \gamma_{c,n}\equiv c  k_{n}\ .
\end{equation}
These equations can be written more compactly using the following
matrices:
\begin{eqnarray}
&&\B.A^{+1}\equiv \left(\begin{array}{cc}1&0\\ 0&1\end{array}\right)\ ,
\qquad \B.A^{-1}\equiv \left(\begin{array}{cc}0&-1\\ 1&0\end{array}\right)\ ,
\nonumber\\
&&\B.C^{+1}\equiv \left(\begin{array}{cc}-1&0\\ 0&1\end{array}\right)\ ,
\quad \B.C^{-1}\equiv \left(\begin{array}{cc}0&1\\ 1&0\end{array}\right) \,,
\end{eqnarray}
which also may be written as $A^{\sigma}_{\sigma'\sigma''}$ and $C
^{\sigma}_{\sigma'\sigma''} $ with $\sigma\,,\s'\,,\s''=\pm 1$.  The
first subscript $\s'$ denotes line ($+1$ corresponds to the upper
line), second subscript $\s''$ denotes column ($+1$ corresponds to the
left one).  Clearly 
\begin{equation}\label{symmetry}
A^{\sigma}_{\sigma'\sigma''}=
A^{\sigma'}_{\sigma\sigma''}\,, \qquad  
A^\sigma_{\sigma'\sigma''}=C^{\sigma'}_{\sigma''\sigma}\ .
\end{equation}
Denoting $u_{n,-1}\equiv u'_n$, $u_{n,+1}\equiv u''_n$ we rewrite
Eqs.~(\ref{sabrareal},\,\ref{sabraima}) as
\begin{eqnarray}
&& {du_{n,\sigma}\over dt}=\Big[ A^\sigma_{\sigma'\s''}\big(\gamma_{a,n+1}
u_{n+1,\sigma'}u_{n+2,\s''}\nl && \quad 
+\gamma_{b,n}
u_{n-1,\sigma'}u_{n+1,\sigma''}\big)+C^\sigma_{\sigma'\sigma''} \gamma_{c,n-1}
 u_{n-2,\sigma'}u_{n-1,\sigma''}\Big] \nonumber\\&&\quad 
-\nu k_n^2~u_{n,\sigma}+f_{n,\sigma}
\ . \label{sabrasig}
\end{eqnarray}
As usual, we adopt the convention that repeated dummy indices (here
$\sigma'$ and $\sigma''$) are summed upon.

Next consider $N$ copies of Eq.~(\ref{sabrasig}). The copies are
indexed by $i,j$ or $\ell$, and these indices take on values
$-J,\dots,+J$, $2J+1=N$, $N$ odd. The $i$th copy is denoted as
$u^{[i]}_{n,\sigma}$.  Let $D^{[ij\ell]}$ be the coupling between
copies, which will be chosen later. Equations~(\ref{sabrasig}) for a
collection of copies are written as
\begin{eqnarray}
&&{du^{[i]}_{n,\sigma}\over dt}=\sum_{j\ell}D^{[ij\ell]}
\Big[A^\sigma_{\sigma'\sigma''}\Big(\gamma_{a,n+1}
u^{[j]}_{n+1,\sigma'}u^{[\ell]}_{n+2,\sigma''}\nonumber\\
&&\quad +\gamma_{b,n}
u^{[j]}_{n-1,\sigma'}u^{[\ell]}_{n+1,\sigma''}\Big)
+C^\sigma_{\sigma'\sigma''} \gamma_{c,n-1}
 u^{[j]}_{n-2,\sigma'}u^{[\ell]}_{n-1,\sigma''}\Big]\nonumber
\\&&\quad  -
\nu k_n^2~u^{[i]}_{n,\sigma}+f^{[i]}_{n,\sigma}
\ . \label{sabracopy}
\end{eqnarray}
To proceed we note that the index $\ell$ is defined modulo $N$, and
introduce a Fourier transform in the ``copy" space:
\begin{equation}
u^\alpha_{n,\sigma}={1\over \sqrt{N}}\sum_{\ell=-J}^J
u^{[\ell]}_{n,\sigma}\exp \Big ({2i\pi\alpha\ell\over N}\Big )
\ . \label{collective}
\end{equation}
Note that the index $\alpha$ is also defined modulo $N=2J+1$.  It is
convenient to consider values $\alpha$ within ``the first Brillouin
zone'' , i.e from $-J$ to $J$.  We will refer to it as the
$\alpha$-momentum. Since $u^{[i]} _{n,\sigma}$ is real,
\begin{equation}\label{sim2}
u^{-\alpha}_{n,\sigma}=(u^{\alpha }_{n,\sigma})^*\equiv u^{\alpha \, *}_{n,\sigma}\ .
\end{equation}
In ``$\alpha$-Fourier space'' Eqs.~(\ref{sabracopy}) read
\begin{eqnarray}
&&\!\!\!\!\!\!\!  {du^\alpha_{n,\sigma}\over dt}= \sum_{\beta,\g} 
\Phi^{\alpha,\beta,\g}[\D_{\a,\b+\g} +\D_{\a+N,\b+\g}  +
\D_{\a,\b+\g+N}]\, \nl
&\times &  \Big\{A^\sigma_{\sigma'\sigma''}
\big[ \gamma_{a,n+1}
u^{\beta }_{n+1,\sigma'}u^{\g}_{n+2,\sigma''}
+\gamma_{b,n}  u^{\beta}_{n-1,\sigma'}u^{\g}_{n+1,\sigma''}
\big] \nl 
&+&
C^\sigma_{\sigma'\sigma''} \gamma_{c,n-1}
 u^\beta_{n-2,\sigma'}u^{\g}_{n-1,\sigma''}\Big\}-
\nu k_n^2~u^\alpha_{n,\sigma}+f^\alpha_{n,\sigma}
\ . \label{sabrafourier}
\end{eqnarray}
where $\D_{\a,\b}$ is the Kronecker symbol: $\D_{\a,\a}=1$ and
$\D_{\a,\b} =0$ for $\a\ne \b$.  Observe that we use Greek indices for
denoting component in $\alpha$-Fourier space, and Latin indices for copies in
the copy space.  As a consequence of the discrete translation symmetry
of the copy index $[i]$ Eqs.~(\ref{sabrafourier}) conserve
$\alpha$-momentum modulo $N$ at the nonlinear vertex, as one can see
explicitly in the above equation.  The coupling amplitudes $\Phi
^{\alpha ,\beta,\gamma}$ in these equations are the Fourier transforms
of the coupling amplitudes $D^{[ij\ell]}$. Our choice of these
amplitudes will be presented in the next subsection.
%%%%%%%%%%%%%%%%%%%%%%%%%%%%%%%%%%%%%%%
\subsection{Choice of coupling}
\label{ss:coupling}%\vskip -.5cm \rightline{\sf ss:coupling }\vskip   .5 cm 
We choose the coupling amplitudes according to
\begin{equation}\label{choice}
\Phi^{\alpha,\beta,\gamma}={1\over
\sqrt{N}}\big [\epsilon+\sqrt{1-\epsilon^2}\, \Psi^{\alpha,\beta,\gamma}\big ] 
\end{equation}
where $\Psi^{\alpha,\beta,\gamma}$ are quenched random phases chosen
with uniform distribution of the phase, independently distributed with
zero average.  The meaning of quenched randomness in the context of a
direct numerical simulation is that we run Eq.~(\ref{sabrafourier})
with a given random choice of $\Phi$, obtain results averaged over the
randomness of the forcing $f$, and then rerun with a fresh random
choice of $\Phi$. Only at the end we average the statistical objects
over the runs. It is believed \cite{98Eyi}, and checked numerically in
this work, that when the number $N\to \infty$ the statistical
functions are self averaging, and the last average is unneeded.

The couplings satisfy the following symmetries:
\begin{eqnarray}
\Psi^{\alpha,\beta,\gamma}&=&\Psi^{\alpha,\gamma,\beta}\,, \qquad 
\left(\Psi^{\alpha,\beta,\gamma}\right)^*=\Psi^{-\alpha,-\beta,-\gamma}\,,
\nl
\Psi^{\alpha,\beta,\gamma}&=&\Psi^{-\gamma,\beta,-\alpha} \ .
\label{conditions}
\end{eqnarray}
The first of these conditions stems from the identity of copies,
leading to the invariance of the equations of motion to an interchange of
copies in the nonlinear term.  The second one is the reality
conditions, the third imposes energy conservation in the inviscid,
unforced limit.  The requirement
\begin{equation}\label{requirment}
\Psi^{\alpha,\beta,\gamma}=1 \quad {\rm if}\quad \alpha\beta\gamma=0 
\end{equation}
guarantees that  for  $N=1$  we recapture  the  original  model at any
$\epsilon$.
Note that for $\epsilon=0$ we have the so called ``Random Coupling
Model" proposed in the context of the Navier-Stokes statistics by
Kraichnan in \cite{61Kra}. It was understood\cite{61Kra,98Eyi} that in
the limit $N\to \infty$ the direct interaction approximation (DIA)
becomes exact. After proper elimination of the sweeping effect (in the
framework on the Lagrangian-history DIA\cite{65Kra} or by means of the
Belinicher-L'vov approach\cite{87BL}) the Kolmogorov 1941 (K41)
scaling appear as an exact solution of the Random Coupling Model.  The
same is true for the shell models\cite{97Pie,98Eyi} in which the
sweeping effect is absent by construction.  For $ \e=1$ the coupling
coefficients in the $\a$-Fourier space ~\Ref.choice ~are
index-independent. This corresponds to uncoupling the
equations~\Ref.sabracopy ~in the copy space, because in this case
$D^{[ij\ell]}=\d_{i,j}\d_{i,\ell}$.  Thus for $ \e=1$ we recover the
original Sabra model with anomalous scaling\cite{98LPPPV}.  We see
that with the choice of the couplings~\Ref.choice ~ the model
interpolates continuously between the ``Random Coupling Model" for
$\epsilon=0$ whose scaling is normal K41 (at $N\to \infty$) and the
Sabra model with anomalous scaling for $\e=1$.  A model of this type
was proposed in the context of the Navier-Stokes statistics by
Kraichnan in \cite{70Kra}. The consequences for the perturbative
theory in the large $N$ and small $\epsilon$ limits were considered by
Eyink \cite{98Eyi}.
%%%%%%%%%%%%%%%%%%%%%%%%%%%%%%%%%%%%%%%%%%
\section{Numerical Investigations}
\label{s:num}%\vskip -.5cm \rightline{\sf ss:num }\vskip   .5 cm 
We are interested in the $N\to \infty$ limit of
Eqs.~(\ref{sabrafourier}) as a function of $\epsilon$. As we discussed
above, for $\epsilon=0$ we have the random coupling model with 
normal K41 scaling in the limit $N\to \infty$, while for $\epsilon=1$ the
copies become uncoupled and with a proper definition of the structure
functions all the scaling exponents $\z_n$ are the same as the scaling
exponent of the original Sabra model. We thus expect that for
intermediate values of $\epsilon$ we may find values of $\zeta_n$ that
interpolate between K41 and the Sabra value.

\begin{figure}
 \epsfxsize=  8.8   truecm
 \epsfbox{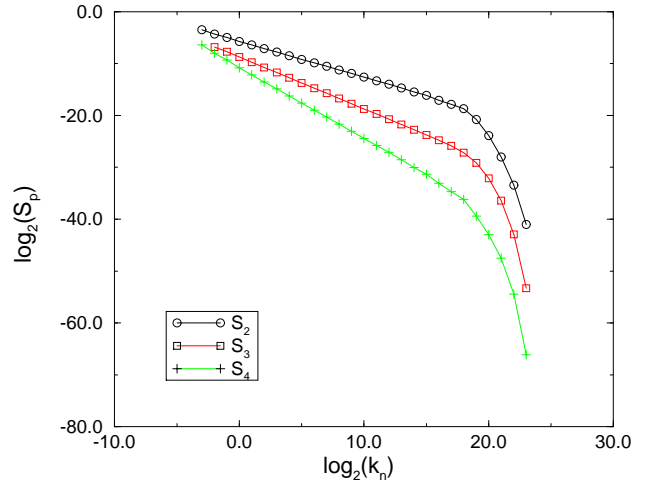}
 \narrowtext
 \caption{Structure functions $\log_2(S_p(k_n))$ 
  vs $k_n$ in  log-log plot for p=2,3,4, $\epsilon=0.8$ and $N=25$.}
 \label{f:num09} 
 \end{figure}

\begin{figure}
\epsfxsize=8.6 truecm
\epsfbox{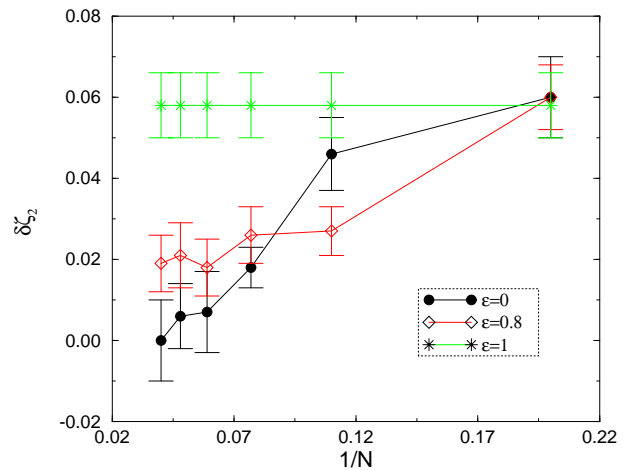}
\narrowtext
\caption{$\delta\zeta_2=\zeta_2-2/3$ vs $1/N$ for $\epsilon=0.8$
(diamonds) and $\epsilon=0$ (circles) for $N$ from 5 to 25.}
\label{f:num08} 
\end{figure}

This is actually what was found.  We measured the scaling exponent 
$\z_2$ of the second order structure
 function $S_2(k_n)\equiv F_2(k_n;0)$, $F_2(k_n;t)$ being defined by
 Eq.~\Ref.F2a . The exponent  $\z_2$  have been calculated either
 with a linear fit in the two decade inertial range, from $n=4$ to
 $n=19$ (see Fig. ~\ref{f:num09}) or by using the fitting
 procedure on the all structure function introduced in \cite{98LPPPV}.
 The results are the same although there is a difference in the
 estimate of the error bars. The error bars in Fig. \ref{f:num08} are
 the ones obtained for the linear fit.
In Fig.~\ref{f:num08} one can see the
plot of the value of the anomalous corrections to Kolmogorov scaling,
$\delta\zeta_2=\zeta_2-2/3$, as function of $1/N$ for $\epsilon=0.8$
together with the same curve for $\epsilon=0$ and for $\epsilon=1$
for $N$ ranging from 5 to 25.

The equations of motion (\ref{sabrafourier}) with $a=1$, $b=c=-0.5$,
were integrated with the slaved Adams-Bashforth algorithm, viscosity
$\nu=4\times 10^{-9}$, a time-step $\Delta t=10^{-5}$.  The forcing
was subjected on the first two shells, chosen random Gaussian with zero
average and with variances such that $\s_2/\s_1=0.7$ (in order to
minimize the input of helicity\cite{98LPPPV}).  Averages were taken
for a time equal to 
%%%%%%%%%%%%%%%%%%%%%%%%%%%%%%%%%%%
\noindent
%%%%%%%%%%%%%%%%%%%%%%%%%%%%%%%
250 eddy turnover times for the case $N=1$. The
averaging times were decreased when the number of copies increased,
taking into account the faster convergence times in these cases.

The random couplings were chosen with a zero-mean, uniform
probability for the random phase at the beginning of each simulations.
Rigorously one should have taken averages over different runs with
different couplings but we checked that self-averaging is already
valid for $N=5$, at least for $\epsilon=0.8$ (very small random
component in the couplings) and within our numerical precision. For
$\e=0$ self-averaging occurs for large numbers of copies.

In Fig.~\ref{f:num08} it is clear then, while in the case for
$\epsilon=0$ the corrections to Kolmogorov scaling goes to zero, 
as was already checked for the equivalent spherical
model \cite{97Pie}, for $\epsilon=0.8$ and for $\epsilon=1$
the corrections converge to a finite value which increases with $\epsilon$.

 %%%%%%%%%%%%%%%%%%%%%%%%%%%%%%%%%%%%
\section{The Equations of Motion of the Statistical Objects}
\label{s:motion}%\vskip -.5cm \rightline{\sf s:motion}\vskip   .5 cm 
\subsection{Multi-time correlation functions: definitions and
  symmetries}
\label{ss:symmetry}%\vskip -.5cm \rightline{\sf ss:symmetry}\vskip   .5 cm 
The nonzero 2nd and 3rd order multi-time correlation functions
of the collective variables are defined as
\begin{eqnarray}
&& F^{\a}_{2;\s,\s'}(k_n;t-t') \D_{\a,\a'}\equiv 
\langle u_{n,\sigma}^\alpha(t) u_{n,\sigma'}^{\a'*}(t')\rangle
\, , \label{F2}\\
&&
F_{3;\s,\s',\s''}^{\a,\a',\a''}(k_n;t,t',t'')\big[\D_{\a+\a',\a''}+\D_{\a+\a',\a''+N}\label{F3}
\\  &&\quad +\D_{\a+\a'+N,\a''} \big] \equiv 
\langle
u_{n-1,\sigma'}^{\alpha}(t) u_{n,\sigma'}^{\a'}(t')
u_{n+1,\sigma''}^{\alpha''*}(t'') \rangle \ .\nonumber  
\end{eqnarray}
The conservation of the $\a$-momentum which was discussed above causes
the 2nd order correlation function to be diagonal in $\a,\, \a'$, and
puts a constraint $\a+\a'=\a''~{\rm mod}N$ on the 3rd order
correlation.  In \Ref.F2 ~we wrote explicitly that in stationary
conditions $F_2$ depends on the time difference $t-t'$ only; it is
clear also that $F_3$ depends on two time differences, say $t-t'$ and
$t-t''$. There are a few nonzero 4th order correlation functions, we
present here a natural general definition:
\begin{eqnarray}
&&
F_{4;\s_1,\s_2,\s_3,\s_4}^{~\a_1,\a_2,\a_3,\a_4}(k_{n_1},
k_{n_2},k_{n_3},k_{n_4};
t_1,t_2,t_3,t_4)\D_{\a_1+\a_2,\a_3+a_4}\nonumber\\
&& \equiv
\langle u_{n_1,\s_1}^{\a_1}(t_1) \, u_{n_2,\s_2}^{\a_2}(t_2)
u_{n_3,\s_3}^{\a_3*}(t_3) \, u_{n_4,\s_4}^{\a_4*}(t_4)
\rangle \ . \label{F4}
\end{eqnarray}

As we see the correlation functions $\B.F_n$ defined by
(\ref{F2})-(\ref{F4}) depend on the indices $\a$ and $\s$.  This
dependence is determined by symmetry considerations. The original
Sabra model \Ref.sabra ~is invariant under the transformation
\Ref.phasesym . This leads to the following symmetry of the
$N,\e$-Sabra model~\Ref.sabracopy :
\begin{equation}
  \label{multisym}
  u_{n,\s }^\a \to \s u_{n,-1}^\a\ .
\end{equation}
Consequently, all the correlation functions must be invariant under this
transformation. For $\B.F_2$ it means $
F^{\a,\a'}_{2;\s,\s'}(k_n;t)=\s\s' F^{\a,\a'}_{2;\s,\s'}(k_n;t)$ and
thus $\s\s' =1$, {\it i.e.} diagonality in $\s,\,\s'$:
\begin{equation}
  \label{diagF2}
  F^\a_{2;\s,\s'}(k_n;t)=\D_{\s,\s'}F^\a_{2,\s}(k_n;t)\ .
\end{equation}
The nonzero components of the $\B.F_3$ tensor are those for which the
product $\s\,\s'\,\s''=1$, namely:
\begin{equation}
  \label{diagF3}
  F^{\a,\a'\a''}_{3;1,1,1}\,,\quad F^{\a,\a'\a''}_{3;1,-1,-1}\,,\quad
  F^{\a,\a'\a''}_{3;-1,1,-1}\,,\quad
  F^{\a,\a'\a''}_{3;-1,-1,1} \ .
\end{equation}
The nonzero components of $\B.F_4$ \Ref.F4 ~are those in which the
product $\s_1\,\s_2\,\s_3\,\s_4=1$. The corresponding list is obvious.

The symmetry of the Sabra model under the phase
transformation~(\ref{phase},\,\ref{theta}) is more restrictive.  For
the ($N,\e$)-Sabra model it leads to an invariance under the
transformation:
\begin{equation}
\left[\begin{array}{c}u_{n,-1}^{\alpha}
                    \\u_{n,+1}^{\alpha} \end{array}\right]
\to
\left[\begin{array}{cc}
   \cos\theta_n,&-\sin\theta_n
\\ \sin\theta_n,&\cos\theta_n
\end{array}\right]
\left[\begin{array}{c}u_{n,-1}^{\alpha}
                    \\u_{n,+1}^{\alpha} \end{array}\right]
\end{equation}
when $\theta_{n}+\theta_{n+1}=\theta_{n+2}$.  The consequence of this
for $\B.F_2$ is rather simple: the only invariant combination of the
2nd order correlation function is the trace
\begin{equation}
F^\a_2(k_n;t) \equiv 
 \sum_{\s=\pm 1}F^\a_{2,\s}(k_n;t)\ .
\label{moduloF2}
\end{equation} 
Actually we can make an even stronger statement: the action of the
random force on the first two shells results in the randomization of
the phases $\theta_1$ and $\theta_2$. Since the other phases are
determined, they are also random, and satisfy
$\left<\cos^2{\theta_n}\right>=\left<\sin^2{\theta_n}\right>=1/2$.
Therefore $F^{\a}_{2,\s}(k_n;t)=\case{1}{2}F^{\a}_2(k_n;t)$.

Moreover, the function $F^{\a}_2(k_n;t)$ is $\a$-independent
\begin{equation}
F^{\a}_{2,\s}(k_n;t)=\case{1}{2}F_2(k_n;t)\ . 
\label{moduloF2/2}
\end{equation} 
This is equivalent to the statement that the correlation functions are
diagonal in the copy space:
\begin{equation}
  \label{diagonal}
  \la u_{n,\s}^{[\ell]}u_{n,\s}^{[\ell']*}\ra \propto \D_{\ell,\ell'}\,,
\end{equation}
To see this note that the deterministic part of coupling in
Eq.~\Ref.sabracopy ~is diagonal in the copy space, {\em i.e.} $\propto
\e \D_{i,j}\D_{i,\ell}$.  The ``inter-copy'' part of coupling (which is
$\propto \sqrt{1-\e^2}$) is random (being a Fourier sum of the
random phases $\Psi ^{\a,\b,\g}$). Therefore every 
realization of $\Psi$'s implies different phase relationships between 
the variables $u_{n,\s}^{[\ell]}$ with different values of $\ell$.
Therefore the only second-order correlation functions 
that survive the $\Psi$-ensemble averaging are those that are
diagonal in $\ell$, $\ell'$. 

Finally we conclude that among the group~\Ref.F2
~of 2nd order correlation function there exist only one object,
$F_2(k_n;t)$, which is invariant under all the symmetry transformations 
of the model and is
independent of the indexes $\a$ and $\s$. In terms of~\Ref.F2 ~it may
be written as:
\begin{equation}
  \label{F2a}
  F_2(k_n;t)= {1\over N}\sum_{\a,\s}F_{2;\s,\s}^{\a }(k_n;t)\ .
\end{equation}
In a similar but much more involved manner one may find the unique
invariant object among the group~\Ref.F3 ~of 3rd  order
correlation functions:
\begin{eqnarray}
  &&F_3(k_n;t,t',t'') 
={1\over N} \sum_{a, \beta,\g} \sum_{\s,\s'\s''}
\Phi^{\g,\b,\a} A^{\s}_{\sigma'\sigma''}[\D_{\a,\b+\g} \nl 
&+&\D_{\a+N,\b+\g}  +
\D_{\a,\b+\g+N}]
F_{3;\s,\s',~\s''}^{~\a,\,\b,\,\g}(k_n;t,t',t'')  \ .
\label{F3a}
\end{eqnarray}
The easiest way to derive this expression is to compute a time
derivative of the invariant object $F_2$~\Ref.F2a ~as is done in
Appendix~\ref{aa:hierF}.  The only objects which may appear in the result
are invariant combinations of the 3rd order correlation functions and
this is exactly the combination~\Ref.F3a . Computing a time
derivative of $F_3$~\Ref.F3a ~ we find the following invariant
combination of the 4th order group~\Ref.F4 :
\begin{eqnarray}
&& F_4(k_n;t_1,t_2,t_3,t_4) \equiv {1\over
  N^2}
\sum_{\a,\b}\sum_{\s_1,\cdots \s_5}A^{\sigma_1}_{\sigma_2\sigma_3}
A^{\sigma_1}_{\sigma_4\sigma_5}
\nl
&\times &
F_{4;~\s_2,\s_3,\s_4,\s_1}^{~~\a,~\b,~\a,~\b}(k_{n+1},k_n,k_{n+1},k_n;
t_2,t_3,t_4,t_1)\ .
\label{F4a}
\end{eqnarray}
There are few other invariant combinations of the 4th order
group~\Ref.F4 , but only the combination~\Ref.F4a   ~appears in
the dynamics.

Note that the normalization constants $N^{-1}$ in
the definitions~(\ref{F2a},\,\ref{F3a}) and $N^{-2}$ in~\Ref.F4a ~are
chosen such that $F_2$, $F_3$ and $F_4$ are of $O(1)$.  Moreover note
that although the triple correlators~\Ref.F3 ~are complex functions
the invariant combination $F_3$~\Ref.F3a ~ is  real as a
consequence of the reality condition for $u^{\alpha}$ and of the
sums over $\alpha$ and $\beta$.  In addition, all these
definitions of correlation functions coincide with the
original correlations of the Sabra model when $N=1$.
%%%%%%%%%%%%%%%%%%%%%%%%%%%%%%%%%%%%%%%%%%%%
\subsection{Hierarchical equations}
\label{ss:hier}%\vskip -.5cm \rightline{\sf ss:hier}\vskip   .5 cm 
We now present the hierarchy of evolution equations for correlation
functions up to the equation for $F_3$, which contains $F_4$ on its
RHS. We will close the hierarchy by expressing $F_4$ in terms of lower
order objects, showing that what we neglect is of higher order in
$\epsilon$.

With the definitions given above, in the inviscid unforced limit ({\sl
  i.e.}  in the bulk of the inertial range)  the evolution equations
derived in Appendix~\ref{aa:hierF} take the form:
\end{multicols}
\vskip -.42cm
\widetext
\leftline{----------------------------------------------------------------------------}
\begin{eqnarray}\label{hierF3} 
{\partial \over \partial t}F_2(k_n,t)
&=& \gamma_{a,n+1}F_3(k_{n+1};0,t,t)
+ \gamma_{b,n}F_3(k_{n};t,0,t)
+ \gamma_{c,n-1}
F_3(k_{n-1};t,t,0)\, ,
\\
 {\partial \over \partial t_1}
F_3(k_n,t_1,t_2,t_3)&=&
\gamma_{a,n}
F_4(k_{n},k_{n+1},k_{n},k_{n+1};t_1,t_1,t_2,t_3)\nl 
&&+ \gamma_{b,n-1}
F_4 (k_{n-2},k_{n},k_{n},k_{n+1};t_1,t_1,t_2,t_3)
+ \gamma_{c,n-2}
F_4 (k_{n-3},k_{n-2},k_{n},k_{n+1};t_1,t_1,t_2,t_3)\,,  \label{hierF41}
\\
 {\partial \over \partial t_2}
F_3(k_n,t_1,t_2,t_3)
&=&
\gamma_{a,n+1}
F_4(k_{n-1},k_{n+1},k_{n+2},k_{n+1};t_1,t_2,t_2,t_3)
\nonumber\\
&&+ \gamma_{b,n}
F_4 (k_{n-1},k_{n-1},k_{n+1},k_{n+1};t_1,t_2,t_2,t_3)
+ \gamma_{c,n-1}
F_4 (k_{n-1},k_{n-2},k_{n-1},k_{n+1};t_1,t_2,t_2,t_3)\,,  
\nl
{\partial \over \partial t_3}
F_3(k_n,t_1,t_2,t_3)& =&
\gamma_{a,n+2}
F_4(k_{n-1},k_{n},k_{n+2},k_{n+3};t_1,t_2,t_3,t_3)
\nonumber  \nl
&&+ \gamma_{b,n+1}
F_4 (k_{n-1},k_{n},k_{n},k_{n+2};t_1,t_2,t_3,t_3)
+ \gamma_{c,n}
F_4 (k_{n-1},k_{n},k_{n},k_{n-1};t_1,t_2,t_3,t_3)\,,  
\nonumber
\end{eqnarray}
\rightline{--------------------------------------------------------------------------}
\vskip -.3cm
%%%%%%%%%%%%%%%%%%%%%%%%%%%%%

\begin{multicols}{2}
This set of equations exhibits terms on the RHS coming from non-linear
contributions only. The linear dissipative terms are negligible in the
limit $\nu\to 0$: there is nothing in the dissipative terms that
compensates the vanishing of the viscosity. In addition, the terms
coming from the forcing appear explicitly only in the equations for
the correlations calculated in $k_1$ and $k_2$, the forced shells.
They vanish in the inertial range.

\begin{figure}\bbox{\hskip -.4cm 
\epsfysize=6.2 truecm
\epsfbox{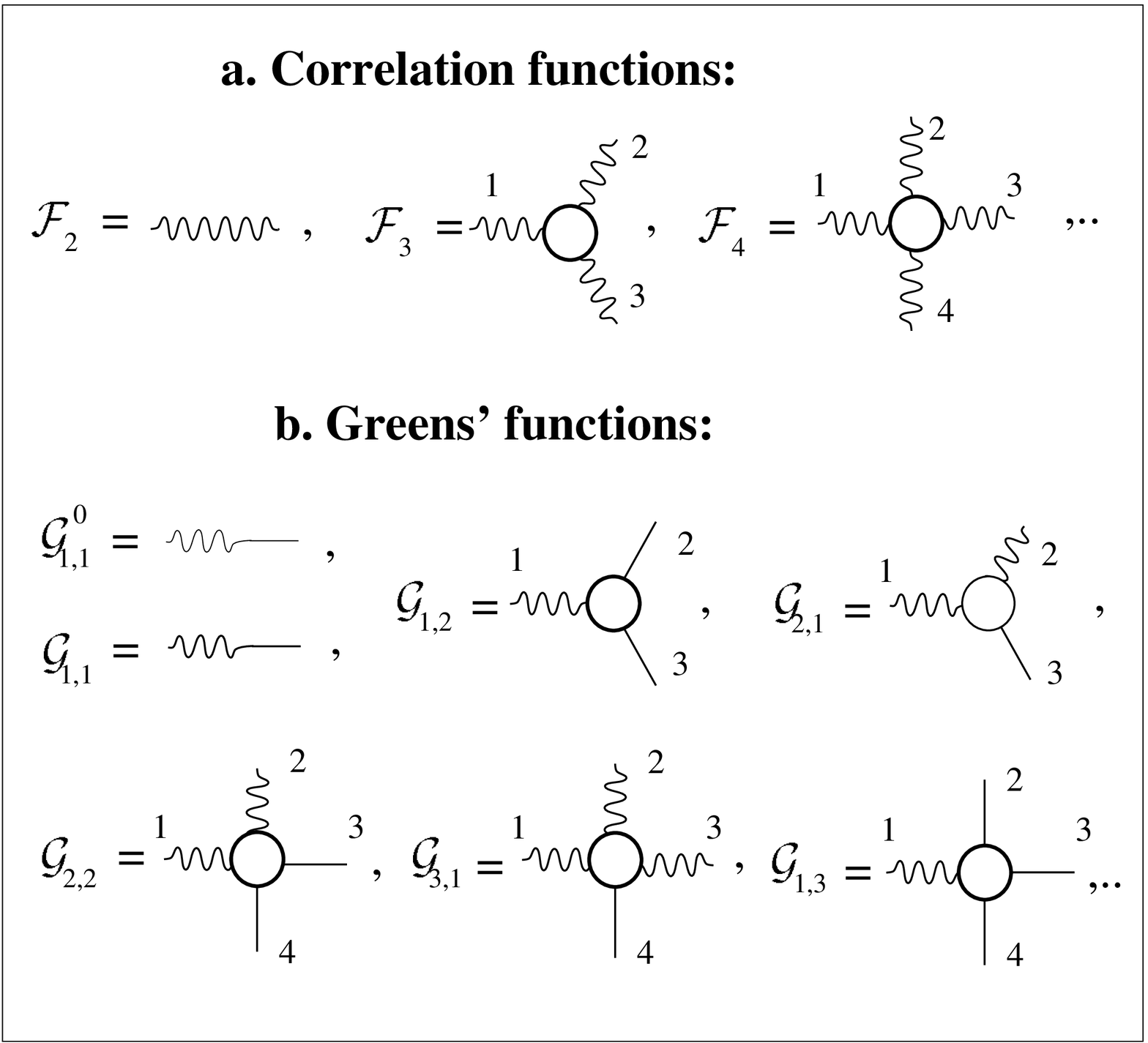}
\epsfysize=6.2 truecm
\epsfbox{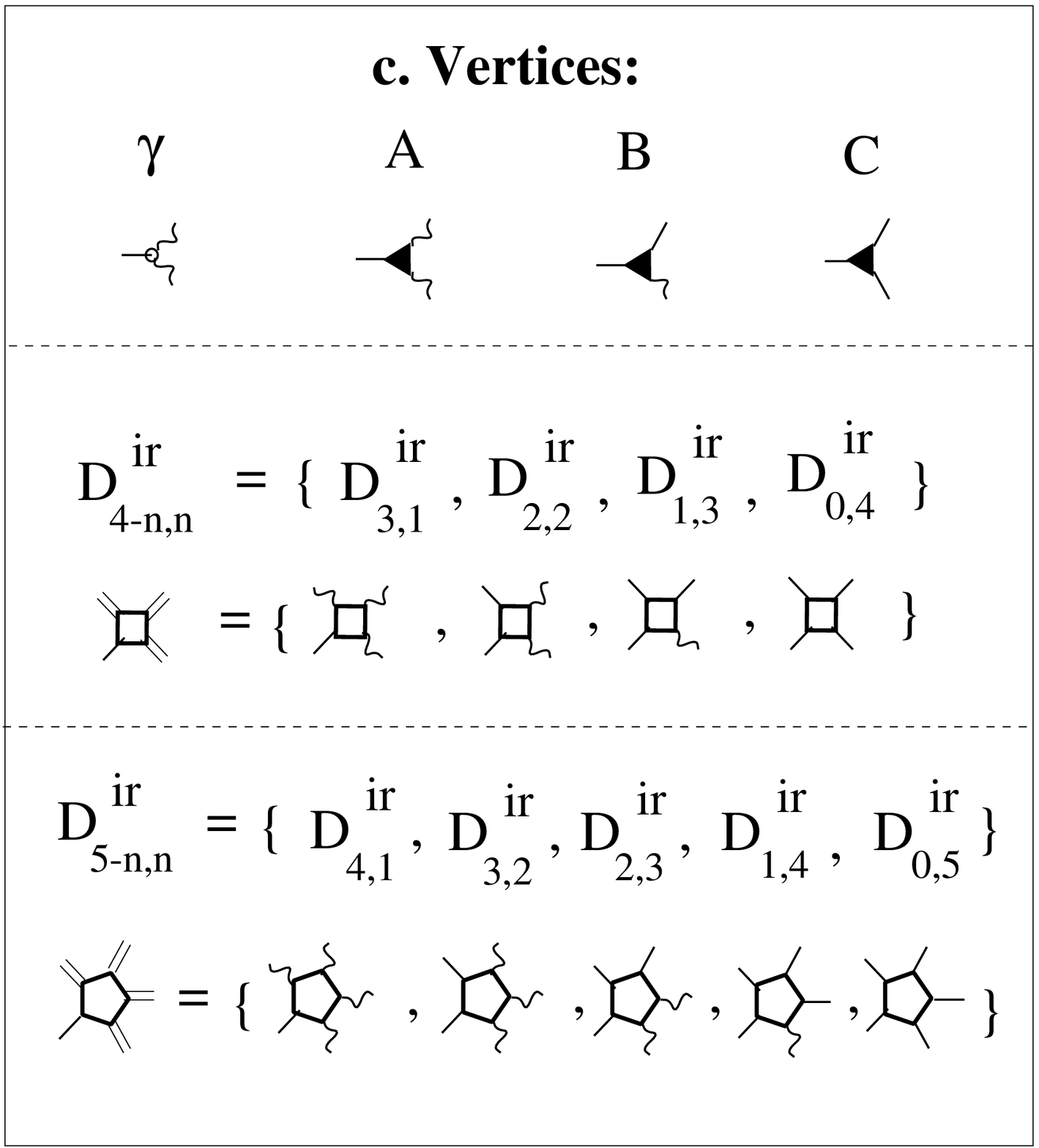} }
\vskip  -6.4 cm~ \hskip 12.1 truecm 
\begin{minipage}{5.25 cm}
\caption{The diagrammatic notation of the basic objects, Panel~a:
  The correlation functions ${\cal F}_n$. Panel~b: the bare Green's
  function (thin line) and the dressed Green's functions ${\cal
    G}_{n,m}$.  A circle connecting $n$ wavy lines stands for an $n$th
  order correlation function, and a circle with $n$ wavy lines and $m$
  straight lines stands for a Green's function with $n$ velocities and
  $m$ forcing.  Panel c : the 3rd order vertices $\gamma$, and $A$,
  $B$, $C$; irreducible contributions of the 4th and 5th order
  vertices $D_{4-n,n}^{\rm ir}$ and $D_{5-n,n}$ denoted as empty
  squares and pentagons.  Double line tails stands either for a
  straight or wavy tail.}
\end{minipage}
\label{f:notation} 
\end{figure}

The structure of the hierarchical Eqs.~\Ref.hierF3 --\Ref.hierF41 ~and
of various future equations becomes transparent if we adopt a
graphical representation. The symbols used in the graphical
representation are those introduced in the context of renormalized
perturbation theory \cite{98LP,multi-sca}, see Fig.~\ref{f:notation}.
It is important to stress however that the present approach is not
perturbative, and does not generate any infinite series of diagrams.
Accordingly the present theory does not suffer from the usual
perturbative problems of unknown convergence properties of infinite
series.

The hierarchical Eqs.~\Ref.hierF3 --\Ref.hierF41 ~ are exhibited in
Fig.~\ref{f:hier}a.  The consistent closure procedure (see below)
requires the introduction of response functions and their equations of
motion.  The equations for all the response functions up to 4-point
objects are presented analytically in Appendix~\ref{a:eq-Green's} and
graphically in Fig.~\ref{f:hier}b,c.
\subsection{Scaling invariance and $h$-slices}
%\label{ss:h-slice}\vskip -.5cm \rightline{\sf ss:h-slice}\vskip   .5 cm 
The Sabra and $(N,\e)$-Sabra equations \Ref.sabra ~ and \Ref.sabracopy
~in the unforced, inviscid limit are invariant under the
following 2-parameter rescaling group $\C.R(h,\r)$:
\begin{eqnarray}
\C.R(h,\r) k_n &=& k_n/\rho\,, \qquad \C.R(h,\r) t=\rho^{1-h}t\,,\nl
\C.R(h,\r) u_n &=& \r^h u_n\ , 
\label{resgr1}
\end{eqnarray}
where $\C.R$ is the rescaling operator.
Correspondingly the hierarchical equations for the correlation
functions and the Green's functions display invariance under the
same rescaling group $\C.R(h,\r)$ when the statistical objects
transform as follows:
\begin{eqnarray}
\C.R(h,\r)  F_n  &=&  \rho^{nh}F_n  \,,  \qquad 
\C.R(h,\r)  G_{n,1}  =  \rho^{(n-1)h-3}G_{n,1}  \,, \nl
   \C.R(h,\r)  G_{1,2}  &=& \rho^{-(h+6)}G_{1,2}  
\,, \dots   \label{resgr2}
\end{eqnarray}
%%%%%%%%%%%%%%%%%%%%%%%%%%%%%%%%%%%
\end{multicols}

%%%%%%%%%%%%%%%%%%%%%%%%%%%%%
\begin{figure}
\bbox{\epsfysize=3.8 truecm \hskip -.2cm \epsfbox{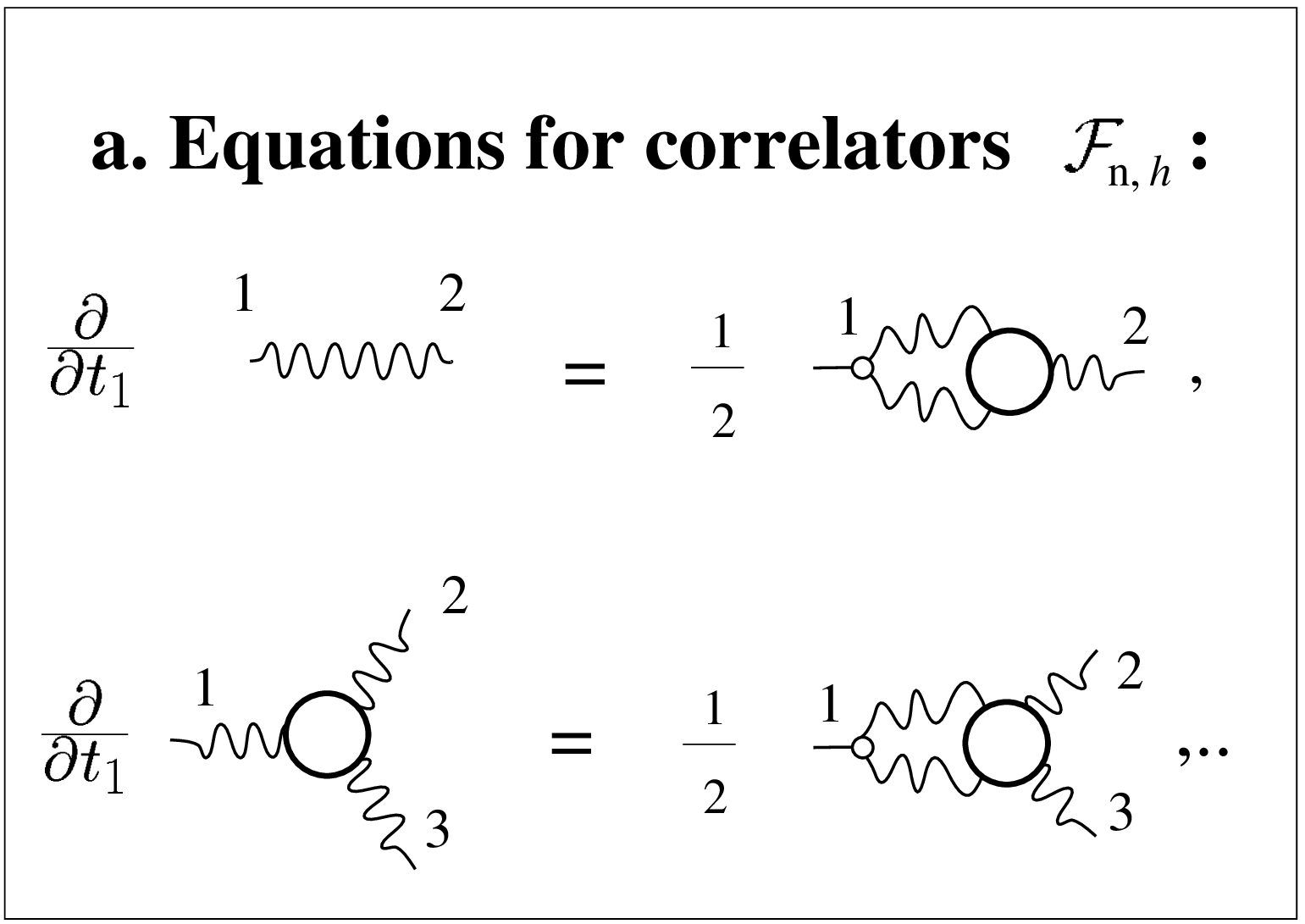} 
\hskip .1  cm  
\epsfysize=3.8  truecm
\epsfbox{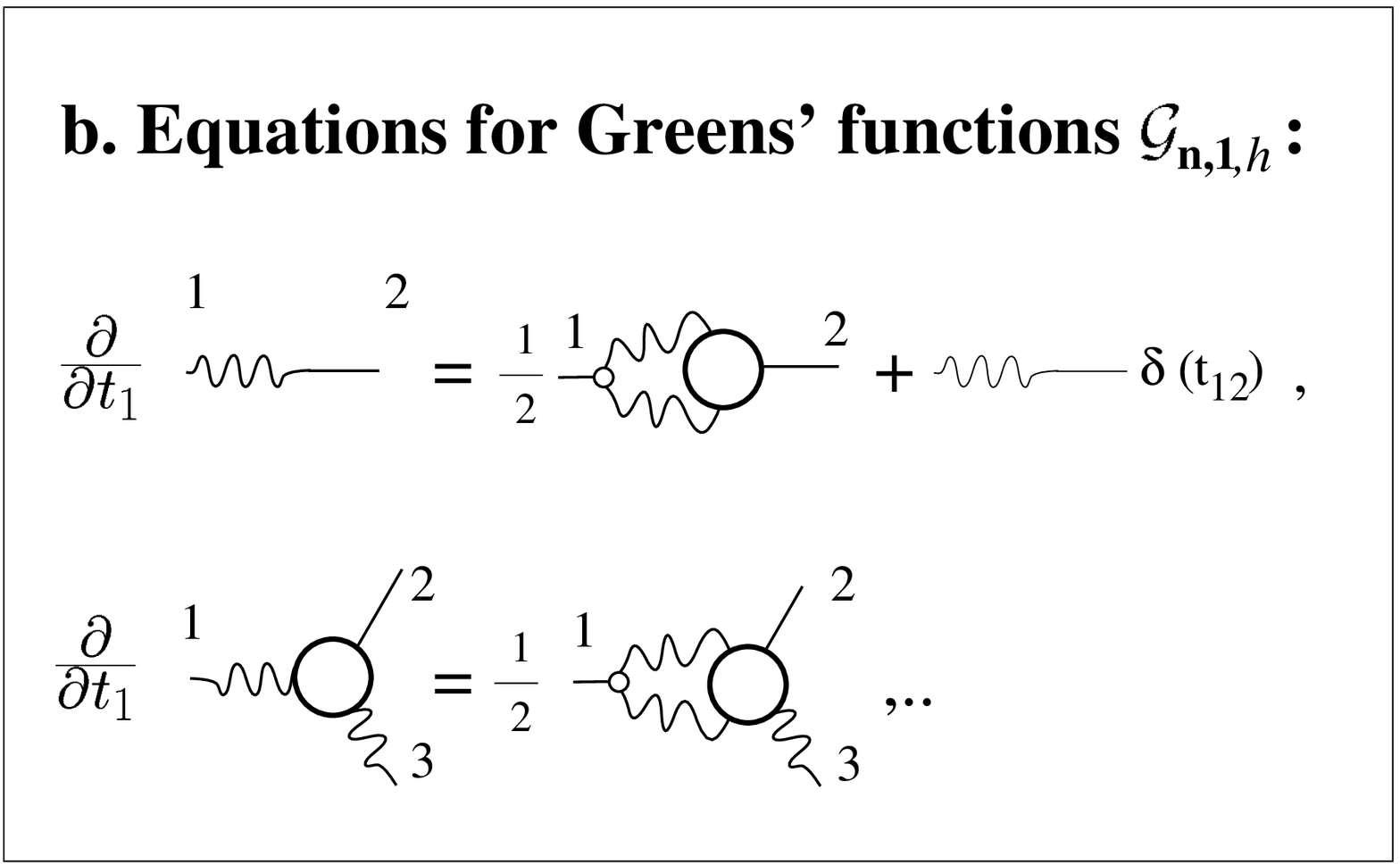}\hskip .1  cm 
\epsfysize=3.8 truecm  \epsfbox{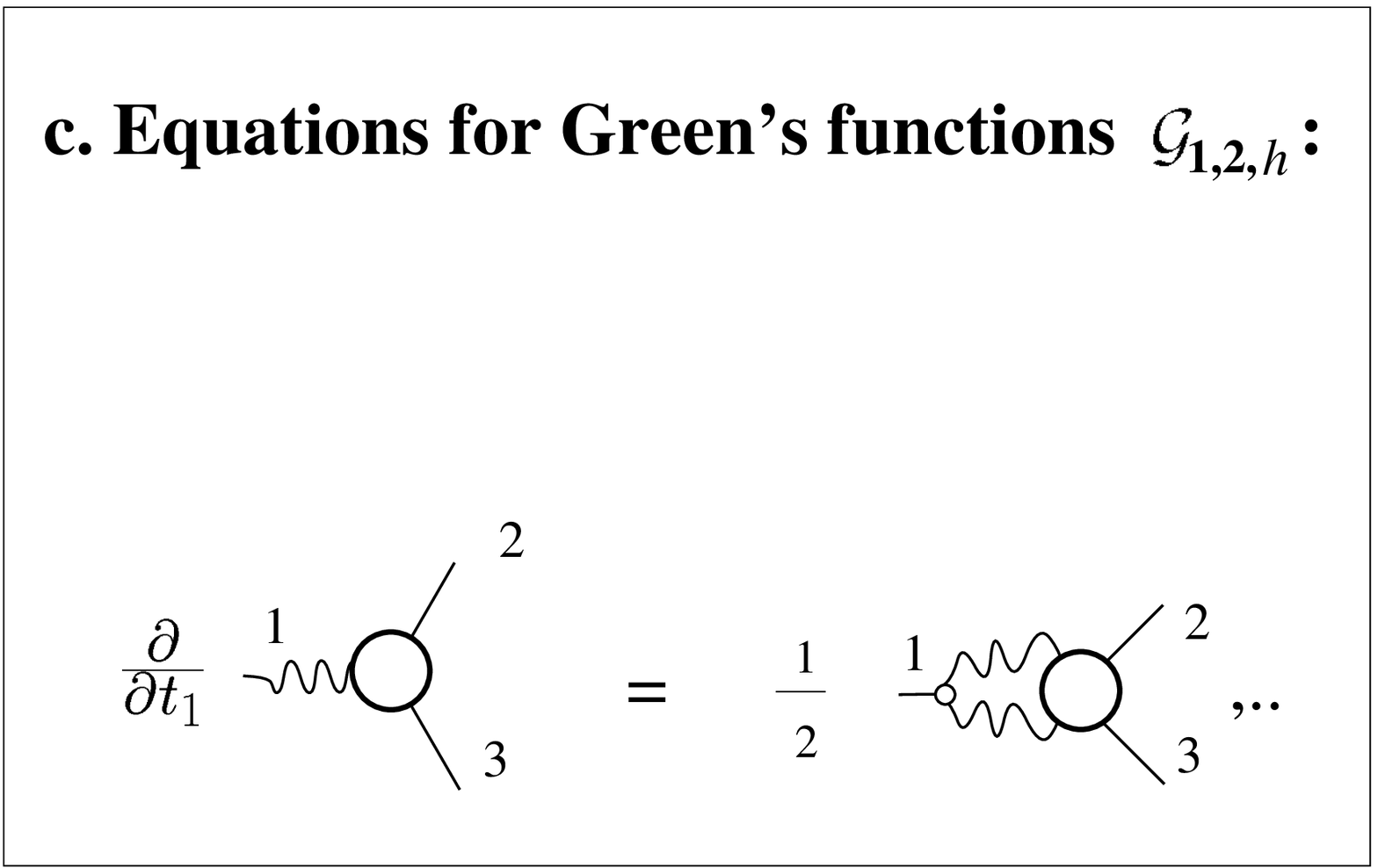}  }
\vskip 0.2cm 
\widetext
\caption{The symbolic representation of the first evolution equations 
  in the hierarchy for correlation and Green's functions.  The
  graphical notations are presented in Fig.~3%\ref{f:notation}
}.  
\label{f:hier}
\end{figure}
%%%%%%%%%%%%%%%%%%%%%%%%%%%%%%
\begin{multicols}{2}

Explicitly, the application of $\C.R(h,\r)$ to any statistical
object reads:
\BEA \label{resgr2a}
&&\C.R(h,\r)  F_n  (k_m,k_{m'},k_{m''},\cdots;t,t',t'',\cdots)\\
&\equiv & F_n  \Big({k_m\over \r},{k_{m'}\over \r},{k_{m''}\over
  \r},\cdots; \r^{1-h}t,\r^{1-h}t',\r^{1-h}t'',\cdots\Big)\ .\nn
\EEA
The simple fact that the hierarchical equations are {\em linear}
equations has a crucial consequence for the theory: by direct
substitution one can check that the rescaling symmetry is {\em wider}
then~\Ref.resgr2  ~and includes an additional $n$-independent function
of $h$ -- ${\cal Z}(h)$, which at present seems arbitrary:
\begin{eqnarray}
\C.R(h,\r)  F_n&=&  \rho^{nh+\C.Z(h)}F_n\,,  \nl
\C.R(h,\r)  G_{n,1}&=&  \rho^{(n-1)h-3+\C.Z(h)}G_{n,1}  \,, \nl
   \C.R(h,\r)  G_{1,2}  &=& \rho^{-(h+6)+\C.Z(h)}G_{1,2}  
\,, \dots   \label{resgr3}
\end{eqnarray}
Note that in principle the rescaling group for the hierarchies of the
correlation and Green's function could be defined using two different
functions ${\cal Z}_1(h)$ and ${\cal Z}_2(h)$. However, the existence
of an inhomogeneous term proportional to the bare Green's function
(first line in Fig.\ref{f:hier}b) results in the identify of ${\cal
  Z}(h)$ in the two hierarchies. This point is explained in detail in
\cite{multi-sca}.

It is tempting to offer a physical explanation of the identity of the
of $\C.Z(h)$. In some way it must be connected with the scaling
exponent of the probability distribution function of subset of
velocity configurations with a given $h$. In other words, if the
probability to find in a turbulent ensemble a solution $\{u_n\}$ of the
unforced equations in the inviscid limit which is $P_h(\{u_n\})$, then
the probability to find a rescaled solution $\C.R(h,\r)\{u_n\} $ is
$\r^{\C.Z(h)}P_h(\C.R(h,\r)\{u_n\})$. It should be stressed however
that the issue of the rescaling properties of the ergodic measure of
our model is far from being exhausted by this comment, and a lot of
further research is needed to solidify and interpret it further.  We
leave this interesting issue for future research.

We reiterate that at this point $\rho$ and $h$ are free parameters,
and ${\cal Z}(h)$ is an $n$-independent arbitrary function of $h$.  We
will see later that this freedom is restricted by the equations:
although $\rho$ remains arbitrary, $h$ belongs to a natural interval,
and the freedom of ${\cal Z}(h)$ is removed by a solvability condition
for the hierarchy of equations.

As stressed before \cite{98LP}, the existence of the rescaling group
$\C.R(h,\r) $ for the linear set of equations suggests the existence
of particular solutions that are characterized by $h$. In other words,
we consider solutions denoted as $F_{n;h}$ and $G_{m,n;h}$ that solve
equations on {\em an $h$-slice}, for example
\begin{eqnarray}
&&{\partial \over \partial t_1}F_{2;h}(k_n,t_1,t_2)
=\gamma_{a,n+1}F_{3;h}(k_{n+1};t_2,t_1,t_1)
\nonumber \\
&& + \gamma_{b,n}
F_{3;h}(k_{n};t_1,t_2,t_1)
+ \gamma_{c,n-1}
F_{3;h}(k_{n-1};t_1,t_1,t_2)\,,
 \label{hierF3h}
\end{eqnarray}
where the rescaling property of a {\em particular} solution $F_{n,h}$ is
\BE\label{prop1} \C.R(h,\r) F_{n,h}=\rho^{nh+{\cal Z}(h)}
F_{n,h}( k_m;t_1,\dots,t_n) \ .  
\EE 
The {\em  general} solution $F_n$ of the hierarchical equations
that we are interested in is naturally obtained as the weighted
integral
\begin{equation}\label{general1}
F_n(\{k_m\},\{t\} )=
\int d\mu(h) F_{n,h}(\{k_m\},\{t\}   ) \,,\label{mf}
\end{equation}
where $\{k_m\}=k_{m'},k_{m''},\cdots $and $\{t\}=t',t'',\cdots$ are the
sets of $k$- and $t$- arguments of correlation functions  $F_n$. 
The weight $d\mu(h)/dh$ may be, in principle, found from the
boundary conditions, by matching the general solution in the inertial
interval with non-universal part of $F_n$ in the energy
containing interval.

The representation~\Ref.general1 ~follows rigorously from the rescaling
symmetry of the problem and the linearity of the hierarchical equations.
It offers a very useful connection to the multi-fractal model for
anomalous scaling in turbulence that was introduced phenomenologically
for simultaneous fully fused objects in \cite{85PF}. In \cite{multi-sca}
it was realized that the multifractal model follows naturally from
the equations of motion for the fully unfused multi-time correlation
functions. Introduce the following dimensionless objects:
\begin{equation}\FL
\tilde F_{n,h}  (\{\k_m\};\{\t\} )\equiv
{1 \over u_0^n } \Big[{\bar k_m \over k_0}\Big] ^{nh+{\cal Z}(h)}
F_{n,h}  (\{k_m\};\{t\} ),
\end{equation}
where $\bar k_m$ is the geometric mean of the $\{k_m\}$, vectors
involved in the correlation function: $k_{m_1}, k_{m_2}\,,\cdots$.
Dimensionless $k$ - and $t$-arguments $\k$ and $\t$ are defined as 
\BE
\label{dym1} \k_{m_j}\equiv { k_{m_j} \over \bar k}\,, \quad
\t_j\equiv {t_j\over t(m_j,h)}\,, 
\EE
where  $t(m,h) $ is a characteristic time on the
$m$-shell for the $h$-slice is 
\BE
 \label{time1}  t(m,h) \equiv
{1\over  k_0 u_0
} \Big( {k_m\over k_0}\Big) ^{1-h}  \ .
\EE
Finally the integral (\ref{mf}) is written as
\begin{eqnarray}
F_n(\{k\},\{t\})
&=& u_0^n\int d\mu(h)
    \Big[{k_0\over \bar k_m }\Big]^{nh+{\cal Z}(h)} 
\tilde F_{n,h}(\{\k\},\{\t\}) \ . \nl \label{mf1}
\end{eqnarray}
It is easy to recognize that this formula is a natural (multi-time,
multi-point) generalization of the multi-fractal representation of
structure functions if ${\cal Z}(h)=3-D(h)$ where $D(h)$ is the
fractal dimension of the set of points that support an H\"older
exponent $h$ \cite{Fri}.  The function $\C.Z(h)$ is related to the
scaling exponents of the structure functions $\z_n$ via the saddle
point requirement
\BE\label{saddle}
\z_n=\min_h[nh+\C.Z(h)]\ .
\EE
This identification stems from the fact that the integral in ~\Ref.mf1
~is computed in the limit $(k_0 /\bar k_m ) \to \infty $ via the 
steepest decent method. Neglecting logarithmic  corrections one find
that $F_n\propto  (k_0 /\bar k_m)^{\z_n}$.

%%%%%%%%%%%%%%%%%%%%%%%%%%%%%%%
\section{${\cal Z}$-covariant Controlled Closure}
\label{s:Zclosure}%\vskip -.5cm \rightline{\sf s:Zclosure}\vskip   .5 cm 
In \cite{multi-sca} we introduced a closure scheme that exploits the
rescaling symmetry of the hierarchy of equation on $h$-slices. This
closure turns out to be somewhat different from traditional closures.
Since the first attempts in the 40's it has been customary to close
the set of equations in the most natural way, that is by expressing
high order correlation functions in terms of lower order ones.
Generally such procedures do not preserve the fundamental rescaling
symmetry (\ref{resgr1}). For example, if one tries to estimate
$F_4$ as $F_2^2$, it would require ${\cal Z}(h)=0$.  
We propose that attempts to close the hierarchy without respecting
the rescaling symmetry (with ${\cal Z}(h)\ne 0$) necessarily lead 
to fundamental mistakes of this type, ruining the possibility of finding 
anomalous exponents.

The minimal requirements for an acceptable closure should be that
realizability conditions are guaranteed: we cannot have negative PDFs
and negative energy spectra. In addition, we want to preserve the
Galilean and rescaling symmetries of the original equations. We therefore 
propose a closure
scheme in which both these requirements are fulfilled.  Realizability
is guaranteed because in our model we have small parameters, i.e.
$1/N$ and $\epsilon$ that are used to control the closure, making the
neglected terms vanishingly small compared with those retained in the
limit $1/N\to 0$ and $\epsilon \to 0$.  The second requirement is met
by expressing higher order correlation (fourth order quantities in the
lowest order closure) in term of lower order ones in such a way that
the rescaling symmetry (\ref{resgr3}) in an $h$-slice is preserved. In
particular all our closure steps are ${\cal Z}(h)$ covariant, in the
sense that all the terms in the resulting equations have the same
${\cal Z}(h)$ dependence.  As a consequence the equations are neutral
to power counting, and ${\cal Z}(h)$ is computed from {\em
  coefficients} rather than powers.  In fact ${\cal Z}(h)$ is found
from solvability conditions as is demonstrated below.

The procedure developed in the rest of this paper is a closure at the 
level of $F_{4;h}$. We are going to represent $F_{4;h}$ in terms of the
lower order quantities, the two point quantities $F_{2;h}$ and $G_{1,1;h}$,
and the three point quantities $F_{3;h}$, $G_{1,2;h}$ and $G_{2,1;h}$. There
is a part of $F_{4;h}$ that {\em cannot} be represented in terms of these
lower order quantities, but we will show that this part is of lower
order in $\epsilon$, in fact of $O(\epsilon^6)$ whereas the retained
terms are of $O(\epsilon^4)$. In order to show this explicitly we
will have to consider the closure at the level of $F_{5;h}$, representing
$F_{5;h}$ in terms of all the lower order 2-point, 3-point and 4-point
objects. 

In representing higher order quantities in terms of lower order
ones there are many different combinations of low order quantities
which appear in any given higher order one. In fact all the topologically
possible combinations are allowed. To facilitate the analysis it is
natural to represent these by diagrams. In making three point objects
from two point ones, there appear three different ways of combining
the two point objects. These are represented by `vertices" which are
denoted below as $A_h$, $B_h$ and $C_h$. In perturbation theories
such vertices are commonly represented in terms of infinite series. In our
approach (as in a fully resumed diagrammatic theory) these vertices are solved
for exactly in terms of the three-point objects on an $h$-slice. 
It turns out, once the theory is written down, that these vertices
appear always in the same combinations with the 2-point Green's functions
$G_{1,1;h}$ which never appears alone. We thus work in terms of three 
newly defined objects that we call ``triplices". The $A$-triplex
is the vertex $A_h$ linked to $G_{1,1;h}$, the $B$-triplex is the
vertex $B_h$ linked to two $G_{1,1;h}$'s, and the $C$-triplex is
the vertex $C_h$ linked to three $G_{1,1;h}$'s. We reiterate that 
at no point do we need to sum infinite series. 

At the end of our procedure we will pose equations of motion for the
$F_{2;h}$ and the three triplices, which are closed upon themselves
and are valid to $O(\epsilon^4)$. A simplified version which
includes only one $A$-triplex is shown in Fig.\ref{f:closed1}.  This
simplified set of equations is analyzed in the sequel, and we show
that as a function of $\epsilon$ it predicts a transition from normal
scaling to anomalous scaling. The reader who is not fascinated by the
technical details can skip Sect.~\ref{s:details}, and observe the
appearance of solvability conditions in Sect.~\ref{s:answer}

\section{Technical Details of Deriving the Closure Equations}
\label{s:details}%\vskip -.5cm \rightline{\sf s:details}\vskip   .5 cm 
\subsection{Definition of the triplex objects $T_{A;h}$, $T_{B;h}$,
$T_{C;h}$}
\label{ss:def-triplex}%\vskip -.5cm \rightline{\sf ss:def-triplex}\vskip   .5 cm 
As a preparatory step for the closure procedure we introduce three
types of triplices: A-triplex $T_{A;h}$, B-triplex $T_{B;h}$
and C-triplex $T_{C;h}$.  Triplices are functions of three $k$- and
three $t$ arguments, say $k$, $k'$, $k''$ and $t$, $t'$, $t''$.  As
any three point objects of the Sabra-shell model they differ from
zero only for three consecutive shell numbers, say $m-1$, $m$ and
$m+1$. The C-triplex is invariant under all permutations of points,
therefore it is enough to
%%%%%%%%%%%%%%%%%%%%%%%%%%%%%%%
\begin{figure}
\epsfxsize=6.truecm \hskip 0.8cm 
\epsfbox{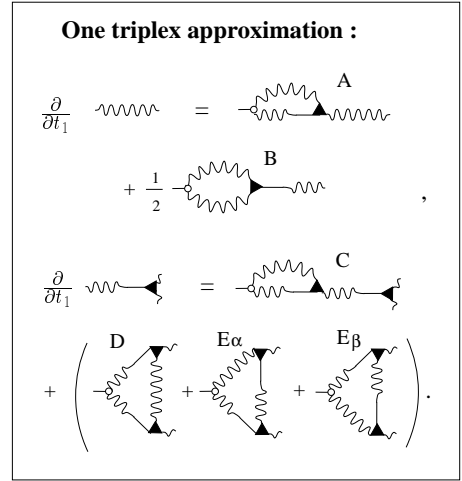}\narrowtext
\caption{
 The hierarchy of equations
  on $h$-slice after the level-4 closure. For the beginning we did not
  consider here contributions of $B$- and $C$- triplices. This allows
  to reduce a system of four equations to the system of just two
  equations for $F_{2;h}$ (two top lines) and for $A$-triplex (two
  bottom lines).    Equation for $A$-triplex follows for equation
in Fig.~8  after disregarding the empty square term and
multiplying the result on the right by two inverse Green's' functions.
\sf f:closed1}
\label{f:closed1}
\end{figure}
%%%%%%%%%%%%%%%%%%%%%%%%%%%%%%%
\noindent
 mention only one of the $k$-arguments,
corresponding, for example to the middle shell $k_m$. So, we have
$T_{C;h}=T_{C;h}(k_m;t,t',t'')$. The convention here is that the first
time argument $t$ corresponds to the smallest shell $m-1$, the second $t'$
to the middle shell $m$ and the last $t''$ to the $m+1$-shell.  The same
convention is chosen for the $A$- and $B$-triplices, the only
difference is that these two are invariant only under permutation of
two points, the third one is special. Therefore we have three
different $A$- and $B$-triplices (and one $C$-triplex) denoted as:
\BE \label{triplex1}
T_{A;h}^{(+1)}\,,\, T_{A;h}^{(-1)}\,,\, T_{A;h}^{(0)}\,;
\quad T_{B;h}^{(+1)}\,,\, T_{B;h}^{(-1)}\,,\ 
T_{B;h}^{(0)}\,;\quad T_{C;h} \ .\EE
The superscript $(+1)$ means here that the $k$-argument of the special
shell is $k_{m+1}$, $(-1)$ corresponds to $k_{m-1}$ and $(0)$ -- to
$k_m$.  These objects are defined in terms of the following seven
three point functions (with the same convention concerning the
arguments, see their definitions \Ref.G21t  --\Ref.G12s :
\BE\label{triplex2}
G_{1,2;h}^{(+1)}\,,\, G_{1,2;h}^{(-1)}\,,\, G_{1,2;h}^{(0)}\,;\quad 
G_{2,1;h}^{(+1)}\,,\, G_{2,1;h}^{(-1)}\,,\, G_{2,1;h}^{(0)}\,;\quad
F_{3;h}\ .
\EE
Three A-triplices $T_{A;h}$ are defined in terms of the three
corresponding (with the same superscript) Green's functions
$G_{1,2;h}$, which describe the response of the velocity $v_{m_1}$ to
two forcing factors on the shells $m_2$ and $m_3$ defined by
Eq.~(\ref{G12s}). In terms of $G_{1,2;h}$ we have:
\end{multicols}
\vskip -.8cm
\leftline{--------------------------------------------------------------------------}
\widetext
\begin{equation}
G_{1,2;h}^{(0)}(k_m,t_1,t_2,t_3)=
\int_{t_1}^\infty dt'_1\int_{t_3}^\infty dt'_3
T_{A;h}^{(0)}(k_m;t'_1,t_2,t'_3)
G_{1,1;h}(k_{m-1};t'_1-t_1)G_{1,1;h}(k_{m+1};t'_3-t_3) \ .
\label{TA0}
\end{equation}
Similar equations serve to define the triplices $T_{A;h}^{+1}$ and
$T_{A;h}^{(-1)}$ simply by replacing on the LHS the appropriate
response functions and by the required changes of the time integration
on the RHS:
\BEA
G_{1,2;h}^{(+1)}(k_m,t_1,t_2,t_3)&=&
\int_{t_1}^\infty dt'_1\int_{t_2}^\infty dt'_2
T_{A;h}^{(+1)}(k_m;t'_1,t'_2,t_3)
G_{1,1;h}(k_{m-1};t'_1-t_1)G_{1,1;h}(k_{m};t'_2-t_2) \,,
\label{TA+}\\
G_{1,2;h}^{(-1)}(k_m,t_1,t_2,t_3)&=&
\int_{t_2}^\infty dt'_2\int_{t_3}^\infty dt'_3
T_{A;h}^{(-1)}(k_m;t_1,t'_2,t'_3)
G_{1,1;h}(k_{m};t'_2-t_2)G_{1,1;h}(k_{m+1};t'_3-t_3) \ .
\label{TA-}
\EEA 
%
%%%%%%%%%%%%%%%%%%%%%%%%%%%%%%%%%%
\vskip -.2cm
\begin{figure}
\bbox{  \hskip  -.3 cm
\epsfysize=6.8 truecm \epsfbox{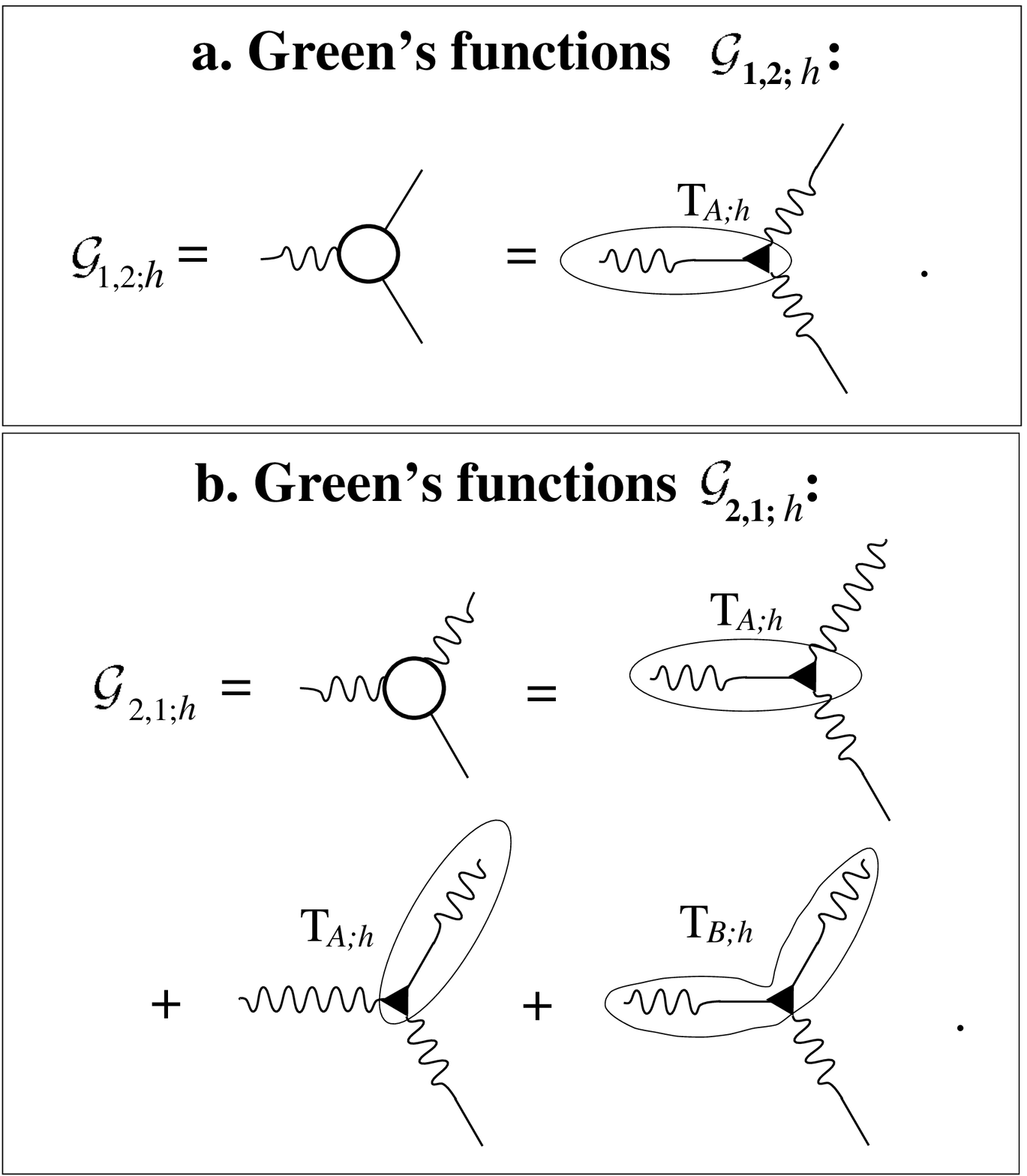} \hskip  .1cm
\epsfysize=6.8  truecm \epsfbox{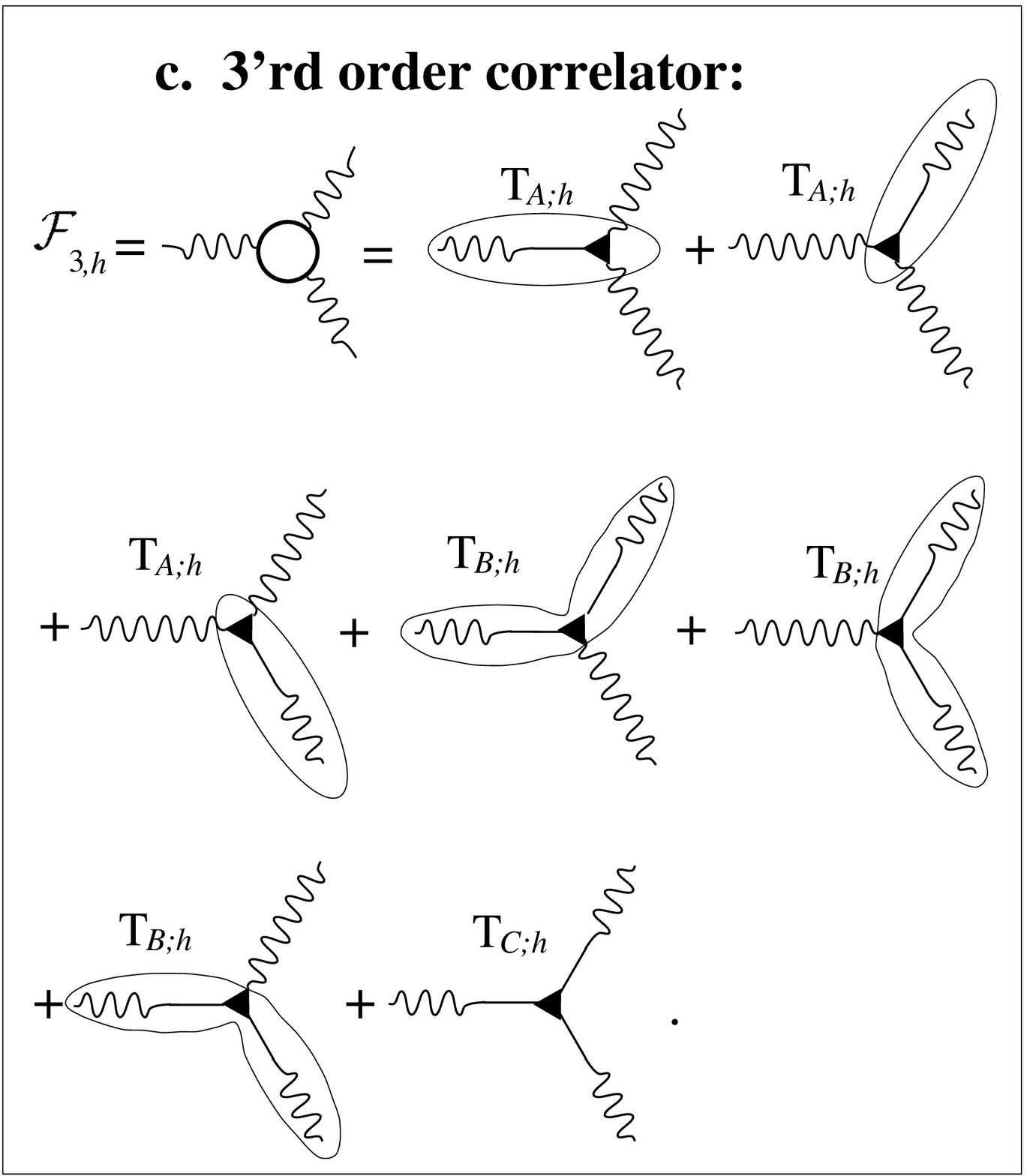} 
}
 \vskip -6.8  truecm ~ \hskip 12   cm 
\begin{minipage}{5.3    cm}
\caption{Exact representation of the Green's functions $G_{2,1;h}$ and 
  $G_{1,2;h}$ (Panels a and  b) and correlation function $F_{3;h}$
 in terms of the vertex $A$ and the Green's' functions
  $G_{1,1}$.  In our approach these serve as the  {\em definitions} of 
  triplices $T_{A;h}$,  $T_{B;h}$, $T_{C;h}$, on an $h$ slice via
  $F_{2;h}$ and $G_{2,1;h}$, $G_{1,2;h}$,  $F_{3;h}$. }
\end{minipage}
\vskip 3cm  
\label{f:exactGF}
\end{figure}
\vskip .6cm \widetext 
Now let us have a look in Fig.~6b at the
representation of the Green's functions $G_{2,1}$ in terms of the two
triplices $A$, $B$, the correlation function $F_2$ and the Green's'
functions $G_{1,1}$. This serves to define the $B$-triplex.  The
explicit analytic definition of $T_{B;h}^{(0)}$ triplex is:
\begin{eqnarray}
&&G_{2,1;h}^{(0)}(k_m,t_1,t_2,t_3)=
\int\limits_{t_2}^\infty dt'_2\Big\{ \int\limits _{0}^\infty dt'_3
T_{A;h}^{(0)}(k_m;t_1,t'_2,t'_3)
F_{2;h}(k_{m+1};t'_3-t_3)
G_{1,1;h}(k_m;t'_2-t_2) + \int\limits_{0}^\infty dt'_1\nl
&\times&
T_{A;h}^{(+1)}(k_n;t'_1,t'_2,t_3)
G_{1,1,;h}  (k_{m};t'_2\!-\!t_2)
F_{2;h}(k_{m-1};t'_1\!-\!t_1) \Big\}  +\int\limits _0^\infty dt'_2
T_{B;h}^{(0)}(k_n;t_1,t'_2,t_3)G_{1,1;h}(k_{m-1};t'_2-t_2)
\label{TB0}
\end{eqnarray}
Similar equations for $T_{B;h}^{(-1)}$ and $T_{B;h}^{+1}$ can be derived
by using $G_{2,1;h}^{(-1)}$ and $G_{2,1;h}^{(+1)}$.  Analogously,
Fig.~6c allows one to define $C$ triples:
\begin{eqnarray}
&&F_{3;h}(k_m,t_1,t_2,t_3)= \int_{0}^\infty dt'_2\int_{0}^\infty dt'_3
T_{A;h}^{(-1)}(k_m;t_1,t'_2,t'_3)
F_{2;h}(k_{m};t'_2-t_2)
F_{2;h}(k_{m+1},t'_3-t_3)\nonumber\\
&&+ \int_{0}^\infty dt'_1\int_{0}^\infty dt'_3
T_{A;h}^{(0)}(k_m;t'_1,t_2,t'_3)
F_{2;h}(k_{m-1},t'_1-t_1)
F_{2;h}(k_{m+1},t'_3-t_3)\nonumber\\
&&+ \int_{0}^\infty dt'_1\int_{0}^\infty dt'_2
T_{A;h}^{(+1)}(k_m;t'_1,t'_2,t_3)F_{2;h}(k_{m-1},t'_1-t_1)
F_{2;h}       (k_{m},t'_2-t_2)\nonumber\\
&& +\int_0^\infty dt'_1
T_{B;h}^{(-1)}(k_m,t'_1,t_2,t_3)F_{2;h}       (k_{m-1},t'_1-t_1)
+\int_0^\infty dt'_2
T_{B;h}^{(0)}(k_m,t_1,t'_2,t_3)F_{2;h}       (k_{m},t'_2-t_2) \nonumber\\
&& +\int_0^\infty dt'_3
T_{B+1,h}^{(0)}(k_m,t_1,t_2,t'_3)
F_{2;h}       (k_{m+1},t'_3-t_3)
+T_{C;h}(k_m;t_1,t_2,t_3) \ .
\label{TC}
\end{eqnarray}
\rightline{--------------------------------------------------------------------------}
\widetext
%%%%%%%%%%%%%%%%%%%%%%%%%%%%%%
\begin{multicols}{2}
\subsection{Representation of the 4-point functions}
\label{ss:rep4}%\vskip -.5cm \rightline{\sf ss:rep4}\vskip   .5 cm 
This theory has four different fundamental $4$-point statistical
objects: the correlation function $F_{4;h}$ and three non-linear
Green's' functions: $G_{3,1;h}$~\Ref.G31 , ~$G_{2,2;h}$ \Ref.G22 , and
~$G_{1,3;h}$.  These objects are shown in Fig.~\ref{f:notation} a as
circles with four tails: four wavy tails for $F_{4;h}$, three wavy and one
straight tail for $G_{3,1;h}$, etc.  As the 3-point objects, also the
4-point objects can be represented as combination of lower order 2 and
3-point objects, and in addition they have a piece that {\em cannot}
be so represented. Consider Fig.7.  For every 4-point
statistical object we show the part obtained from 2-point objects,
which is the Gaussian decomposition, a part that is given in terms of
2-point objects and 3-point vertices, and finally the part that calls
for the introduction of 4-point vertices which are represented by
empty squares and denoted by $D$. There are four different
$D$-vertices, depending on the nature of the tails connecting them.
As in the case of the 3-point vertices, the 4-point vertices that can
be solved for in terms of the 4-point statistical objects. The figures
offer a strong motivation for the graphic representation. Analytically
the equation for $F_{4;h}$ consist of 63 terms, Eqs. for $G_{3,1;h}$
has 31 terms, {\em etc}. The graphic notation is sufficient for all
our future considerations.
\end{multicols}
 \vskip -.5cm
%%%%%%%%%%%%%%%%%%%%%%%%%%%%%%
 \begin{figure}
\epsfxsize=10.5   truecm  ~   \hskip -.55cm
\epsfbox{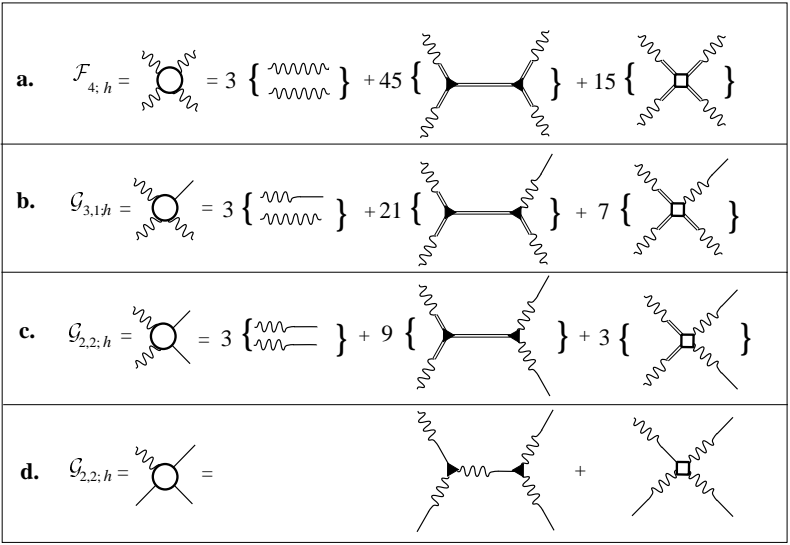}  
\vskip -7.3cm ~\hskip  10.3   truecm
\begin{minipage}{7.2cm}
\caption{Compact representation of the 4th-order statistical objects 
  in terms of triplices and irreducible 4th order vertices
  (empty squares).  Double lines denote either wavy or straight lines.
  Thus a short wavy and short double line denote either pair
  correlator (long wavy line, when the short double line denotes a
  wavy one) or the Green's' function (short wavy and short straight
  line, when the double line denotes a straight one). One understands
  a long double line as consisting of two short double line. Therefore
  a long double line may denote 3 versions of 2-point objects: the
  2-point correlation function and 2 orientations of the Green's
  functions. The long double line is not allowed to be made of two
  short straight lines.  The different topologies are counted in these
  figures, for example, Panel a one has 3 different Gaussian
  decompositions of $F_{4;h}$, 45 different contributions in terms of
  triplices and 15 contributions in terms of 4-point vertices. }
\end{minipage}    
\label{f:compactFG4}
\end{figure}
\begin{multicols}{2}

%%%%%%%%%%%%%%%%%%%%%%%%%%%%%%%%%
In particular, by substituting these representations on the RHS of
the hierarchy of equation for the 3-point objects (Panel c and
bottom lines of the Panels a and b of Fig.~\ref{f:hier}) we get
Equations for the 3-point objects in terms of the 2, 3-point
objects and 4-point vertices. An example of such an equation is
given in Fig.~8.
\subsection{Closure on the level of 4th order objects}
\label{ss:4clos}%\vskip -.5cm \rightline{\sf ss:4clos}\vskip   .5 cm 
The simplest closure which may already predict anomalous scaling is
very simple, and is obtained by neglecting the empty squares in all
the equations of motion for 2- and 3-point objects.  Below we show
that in this approximation we account exactly for all terms up to
$O(\e^4)$ and neglect terms of $O(\e^6) $ and smaller.  The resulting
system of equations can be exactly reduced to a system of four
evolution equations for $F_{2;h}$ and the three triplices. A
simplified version of these equations involving the A-triplex only is
shown in Fig.~\ref{f:closed1}.

%%%%%%%%%%%%%%%%%%%%%%%%%%%%%%%
\begin{figure}\label{f:eqG12}
\epsfxsize=5.5 truecm ~ \hskip -.6cm 
\epsfbox{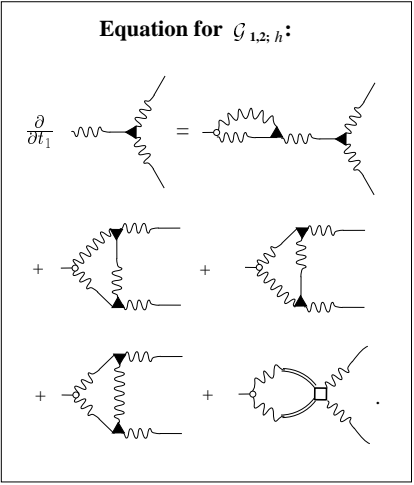}
\vskip -6cm ~\hskip 5.3   cm
\begin{minipage}{3  cm}
\caption{Equation for $G_{1,2;h}$. For simplicity we did not 
  display here the   terms with $B$-  and $C$-triplices. Topologically
  they look like the exhibited terms with the $A$-triplex.   } 
\end{minipage}\vskip 2 cm 
\end{figure}
%%%%%%%%%%%%%%%%%%%%%%%%%%%
\subsection{Closure on the level of 5th order objects}
\label{ss:5clos}%\vskip -.5cm \rightline{\sf ss:5clos}\vskip   .5 cm 
In order to formulate this closure consider four sets of hierarchical
equations:

\noindent $\bullet$
The first set consists of 3 evolution equations for the correlation
functions $F_{2;h}$, $F_{3;h}$ and $F_{4;h}$. These equations contain
on their RHS $F_{3;h}$, $F_{4;h}$ and $F_{5;h}$ respectively. 

\noindent $\bullet$
The second set consists of 3 equations for the Green's' functions
$G_{1,1;h}$, $G_{2,1;h}$, and $G_{3,1;h}$, having on their RHS the
$G_{2,1;h}$, $G_{3,1;h}$, and $G_{4,1;h}$.  

\noindent $\bullet$
The third set consists of 2 equations for the Green's' functions
$G_{1,2;h}$ and $G_{2,2;h}$ in terms of $G_{2,2;h}$ and $G_{3,2;h}$.

\noindent $\bullet$
The fourth set has just one equation for the Green's function
$G_{1,3;h}$ in terms of $G_{2,4;h}$.

The new objects that appear at this level are the four 5th order
functions $F_{5;h}$, $G_{2,3;h}$, $G_{3,2;h}$ and $G_{4,1;h}$ which
appear in the RHS of the last equation in each of the four sets. Like
the 4th-order objects they may be presented as a sum of reducible and
irreducible parts, the latter being related to a new set of 5th order
vertices that can be represented graphically as empty pentagons.  The
reducible part is a sum of all the topologically allowed contributions
made from lower order objects: $F_{2;h}$, $G_{1,1;h}$, three triplices
and four 4th order vertices $D_{4-n,n;h}$. The irreducible part of the
statistical objects is, by definition, what is not reducible.  The
Closure on level of 5th-order objects is obtained by disregarding all
the irreducible contributions in 5th-order objects.

\begin{figure}%%%
\epsfxsize=8truecm \hskip .2cm 
\epsfbox{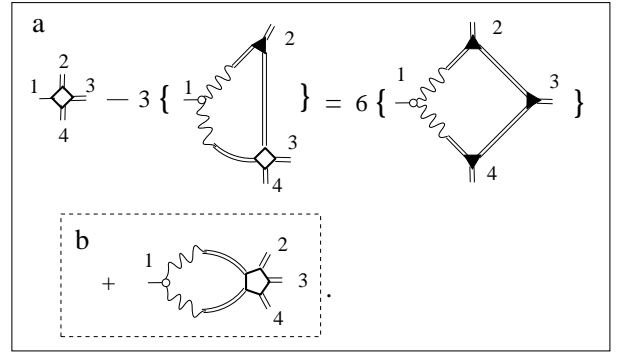}
\narrowtext
\caption
{Exact equation for $D_{n,4-n}^{\rm ir}$ (empty square). }
\label{f:repD4-D5}
\end{figure}

We are not going to study explicitly the equations obtained from the
closure at the level of the 5th order objects. We use this level to
justify the relative smallness of what was neglected in the closure on
the level of the 4th order objects. To this aim we derive now an exact
equation for the empty squares that does not contain any time
derivative. This is done as follows: Compute $\partial
F_{4;h}/\partial t_1$ using the representation of $F_{4;h}$ via
$D_{4-n,n;h}$ and 2nd order objects, shown in Fig.~7.
For the time derivatives of the 2nd order objects use the
corresponding evolution equations involving $F_{3;h}$ and the third
order Green's functions.  The latter in their turn may be expressed
exactly via 3rd order vertices (black triangles).  On the other hand
we have the evolution equation for $\partial F_{4;h}/\partial t_1$
which relates it to $F_{5;h}$.  Now $F_{5;h}$ can be presented as a
sum of reducible contributions (made from lower order objects) and an
irreducible part, which is presented via empty pentagons. Equating the
formulae for $\partial F_{4;h}/\partial t_1$ one gets after some
tedious algebra an exact {\em linear} equation for the 4th order
vertices in terms of lower order objects and the 5th order empty
pentagons. This equation is displayed in Fig.~\ref{f:repD4-D5}. The
closure on the 5th level amounts to discarding the empty pentagons,
leading to a linear inhomogeneous equation for the empty squares. This
equation serves for the order of magnitude estimates presented in the
next section.
%%%%%%%%%%%%%%%%%%%%%%%%%%%%
\noindent
%%%%%%%%%%%%%%%%%%%%%%%%%%%%%%%%%%

\section{Orders of magnitude}
\label{ss:mag}%\vskip -.5cm \rightline{\sf ss:mag}\vskip   .5 cm 
 In this Section we explain how the small parameter $\e$ controls
  the sequential steps in the closure procedure in the $(N,\e)$-model. 
In particular, we address the empty square
  in Fig.~8 that is being discarded in the closure on
  the level of 4th order objects, showing that it is of lower order in $\epsilon$
  compared to the terms that are retained in the limit $N\to
  \infty$.

First, recall that initially all our objects had  ``Fourier
space'' indices $\a$, $\b$, $\g,\dots$ and spinor indices $\s$, $\s'$, $
\s''\dots$. The tensorial correlation functions $F_{2;\s,\s'}^\a$,
$F_{3,\s,\s',\s'' }^{\a,\a',\a''}$ and $F_{4,\s,\s'\dots
  }^{\a,\a'\dots}$ are defined by Eqs.~\Ref.F2 -- \Ref.F4  and the tensorial
Green's functions of the 2nd, 3rd and 4th order are defined by
Eqs.~\Ref.G11t --\Ref.G12t .  Equations \Ref.F2a -- \Ref.F4a ~introduce
the scalar objects $F_{2}$, $F_3$ and $F_4$ as sums over tensor indices.
For these objects we derived scalar equations and that were represented
diagrammatically. Of course,  all the diagrammatic equations (including
equations on an $h$-slice) discussed so far may be understood also as
equations for the tensor objects with explicit $\a$, $\b$, $\g,\dots$ and $\s$,
$\s'$, $\s''\dots$ indices. In this case each equation (or diagram)
{\em for the tensor objects} includes summations over dummy Fourier and
spinor indices for all the intervening objects, (vertices, triplices, {\em etc.}),  
in the body of the diagrams. This calls for definitions of tensorial triplices
and fourth and fifth order vertices. In particular, each 
triplex is understood as a
sum similar to \Ref.F3a ~of objects depending on $\alpha,\beta$'s.
For example:
\BEA&& T_{A;h}^{(0)}(k_{n};t_1,t_2,t_3)=\frac{1}{N} \sum_{a,
  \beta,\g}[\D_{\a,\b+\g} +\D_{\a+N,\b+\g} \nl &+& \D_{\a,\b+\g+N}]
T_{A;h}^{(0)\alpha,\beta,\g}(k_{n};t_1,t_2,t_3)\,,
\label{Tsum} 
\EEA
where the sums on $\sigma$'s have already been performed.  The objects
$T_{A;h}^{\alpha,\beta,\g}$ are obtained from tensorial equations
identical to (\ref{TA0}-\ref{TC}), but with the 2nd and 3rd order
objects having the appropriate Greek indices.  Having done so it
follows that $T_{A;h}^{\alpha,\beta,\g}$ is of order $O(1/\sqrt{N})$:
this stems from the fact that $F^{\alpha,\b,\g }_{3;h}$ is of order
$O(1/\sqrt{N})$ and $F^{\alpha}_{2;h}$ and $G^{\alpha\alpha}_{1,1;h}$
are $O(1)$.  In other words, $T_{A;h}^{\alpha,\beta,\g}$ has the same
order magnitude {\em and the same tensorial structure} as the 3-point
correlation function $F_{3;h}^{\alpha,\beta,\g}$. Again,
Eq.~(\ref{Tsum}) can be inverted to give:
\begin{equation}
T_{A;h}^{(0)\alpha,\beta,\g}=
\Phi^{-\alpha,-\beta,-\g}T_{A;h}^{(0)} \ , \label{Tphi}
\end{equation}
where $ T_{A;h}^{(0)}$ is a function $O(1)$. Similar relationships may
be written for the other triplices.

Using this information we can discuss the order of magnitude of the
terms in the equation for $G_{1,2;h}$ exhibited in Fig.~8.
All the following considerations apply similarly to the equations for
$F_{3;h}$ and $G_{2,1;h}$. In fact the order of magnitude of the
different terms in the equations does depend only on the topology of
the graphs and not on the kind of 2-point functions.

The equation in Fig.~8 has three types of contributions.
The first one on the RHS has a ``simple'' loop (made of  2  lines
representing 2nd order objects). The next three terms have a loop made from 3
lines (``triangular'' loop) and the last term has a 4th order vertex,
an empty square.

The term with the simple loop has two sums over Greek indices, each
having a product of two complex conjugate $\Phi$'s (three factors
$\Phi$'s coming from the three vertices and one coming from the
definition of $G_{1,2;h}$).  Accordingly this term is of the order of
unity, as the LHS.  In the triangular loops  we also have two sums on
Greek indices of four factors of $\Phi$.  In this case the term will
be multiplied by the following factor
\begin{eqnarray}\label{4phi}
&&\!\!\!\! \frac{1}{N}\sum_{\alpha\beta\delta}\Phi^{\alpha+\beta,\alpha,\beta}
\Phi^{\alpha,\delta,\alpha-\delta}
\Phi^{\beta,-\delta,\beta+\delta}
\Phi^{\alpha-\delta,\alpha+\beta,-\beta-\delta}   \\
&=&\frac{1}{N^3}\sum_{\alpha\beta\delta}
(\epsilon+\sqrt{1-\epsilon^2}\Psi^{\alpha+\beta,\alpha,\beta})
(\epsilon+\sqrt{1-\epsilon^2}\Psi^{\alpha,\delta,\alpha-\delta})\nonumber\\ &\times&
(\epsilon+\sqrt{1-\epsilon^2}\Psi^{\beta,-\delta,\beta+\delta})
(\epsilon+\sqrt{1-\epsilon^2}\Psi^{\alpha-\delta,\alpha+\beta,-\beta-\delta})\ 
.\nonumber 
\end{eqnarray}
The four factors $\Psi$ do not appear as couples of complex
conjugates. As a result, after averaging over the randomness and
sending $N\to \infty$ all the terms that present a random phase in
(\ref{4phi}) will go to zero and only the factor $\epsilon^4$ will
survive. The conclusion, which is of fundamental importance for the
solution of our model, is that 
{\it terms with triangular loops are smaller than terms with a simple
loop by a factor of $O(\e^4 )$}. This is the first place that
$\epsilon$ appears as an important parameter that distinguishes the
order of magnitude of competing terms. We reiterate that usually, without
the introduction of $\epsilon$, all the diagrams that one obtains
in the context of turbulent statistics are of the same order.

The second important role of the small parameter $\epsilon$
is to make the last term with an
empty square even smaller than the terms just analyzed. We show that it
is of the order of $\e^6$. In order to see
this, consider the equation for the empty squares exhibited in
Fig.~\ref{f:repD4-D5}.  The irreducible part of $F_{5;h}$ is related here
to the empty pentagon. At this point we discard the empty pentagon
without much ado, and find the order of magnitude of the empty square.
Of course, this step should be justified by estimating next the order
of the empty pentagon, which can be done by going to the closure on
the level of 6th order objects [with the result that the term
with the empty pentagon is smaller than the other terms  in this
equations by a factor of $O(\e^2)$. This term is  of $O(\e^8)$].

After discarding the empty pentagon we have an equation for the empty
square in terms of itself and of lower order objects. This equation
corresponds to the top line in Fig.~\ref{f:repD4-D5}.  On the RHS we
have objects with four vertices in one loop.  Accordingly they are of
$O(\epsilon^4)$. The LHS has an operator that acts on our empty
square. The operator contains unity, and is therefore estimated to be
of the $O(1)$. This is in fact a dangerous point. We are not guaranteed
{\em a priori} that the empty square is not a zero mode
of this operator.  We are going to make an assumption that this is not the case.
This assumption cannot be justified before the end of the calculation.
In principle, one needs to proceed under this assumption, compute explicitly the
2nd and 3rd order objects, and justify the assumption {\em a posteriori}
by a direct substitution and calculation. 

Making this assumption we conclude that the term including the empty
square in Fig.~8  is of $O(\epsilon^6)$ and it can be
neglected with impunity. This leaves us with the equation represented
in Fig.~\ref{f:closed1}.  Previously we were satisfied with this
graphic representation but now we need to discuss its analytic form.
In Appendix~\ref{a:Aclos} we display the algebraic form in the
$\o$-representation in full.  For the scale invariant
functions~\Ref.F2inv ~ these equations may be written as follows:
\BEA
\label{eqF2} 
-[i \O +\tilde\sigma_h( \O )&]&\tilde \C.F_{2;h}( \O)= \tilde
\C.B_h(\O )\,, \\  \label{eqTp} 
\mu [i\Op+\tilde \sigma^*( \O _+) &]&\tilde \C.T_{A;h}^{(+1)}( \O _-,\O_
0)\\ &&=\e^4 \big[ \tilde \C.D_h^{(+1)}(\O_ -,\O_ 0 ) + \tilde
\C.E_h^{(+1)}(\O_ - ,\O_ 0 ) \big] \,, \nl  \label{eqTz} 
 -[i \Oz +\tilde \sigma_h(
\Oz )&]&\tilde \C.T_{A;h}^{(0)}( \Om,\Op)  \\
&&= \e^4 \big[ \tilde \C.D_h^{(0)}(\Om,\Op )+ \tilde
\C.E_h^{(0)}(\Om,\Op ) \big] \,, \nl  \label{eqTn} 
-[i\Om+\tilde \sigma( \O _- )&]& \tilde
\C.T_{A;h}^{(-1)}(\O_0,\O_+)\\ &&=\e^4 \big[ \tilde
\C.D_h^{(-1)}(\O_0,\O_+ ) + \tilde \C.E_h^{(-1)}(\O_0,\O_+ )\big]\ .
\nn \EEA
Here $\tilde \C.F_{2;h}$ is a dimensionless 2nd order correlation
functions, $ \tilde \C.T_{A;h}^{(\ell)}$ are dimensionless triplices.
The term with the mass operator $\tilde \s$ originates from the
diagram A in the equation for $\tilde \C.F_{2;h}$ and from diagram C in
the Eqs.  for $ \tilde \C.T_{A;h}^{(\ell)}$. The terms $\tilde
\C.E_h^{(\ell)}$ and $\tilde \C.D_h^{(\ell)}$, with a factor $\e ^4$
in front, originate from the triangle diagrams D and E in
Fig.~8.  Analytical forms of these functions are given in
Appendix~\ref{a:Aclos}.

%%%%%%%%%%%%%%%%%%%%%%%%%%%%%%%
\section{Preliminary analysis of the solvability conditions}
\label{s:answer}%\vskip -.5cm \rightline{\sf   s:answer}\vskip   .5 cm 
Equations \Ref.eqF2   --\Ref.eqTn , which are exact in the limit $\e\to 0$, are
nonlinear integral equations for one function of one frequency $\tilde
\C.F_{2;h}( \O)$ and three functions $\tilde \C.T_{A;h}^{(\ell)} $ (for
$\ell=0 $ and $\ell=\pm 1$) of two frequencies. A complete analysis of
these equations is very difficult and beyond of the scope of this
paper.  In this section we will only demonstrate how the solvability
conditions of these equations may determine the function $\C.Z(h)$.
For the purpose of this demonstration it is sufficient to restrict
ourselves to the ``{\em one-triplex reduction}" of Eqs.~\Ref.eqF2 --
\Ref.eqTn .  Considering only the two equations for $\tilde \C.F_{2;h}$
and $\tilde \C.T_{A;h}^{(0)} $, we substitute on the RHS of these
equations $\tilde \C.T_{A;h}^{(0)} $ instead of $\tilde
\C.T_{A;h}^{(\pm 1)}$:
\BEA
\label{oneF2} 
-[i \O +\tilde\sigma_h( \O )&]&\tilde \C.F_{2;h}( \O)= \tilde \C.B_h(\O )\,, 
\\ \label{oneTz} 
 -[i \Oz +\tilde \sigma_h(
\Oz )&]&\tilde \C.T_{A;h}^{(0)}( \Om,\Op)  \\
&&= \e^4 \big[ \tilde
\C.E_h^{(0)}(\Om,\Op ) + \tilde \C.D_h^{(0)}(\Om,\Op )\big] \ .
\nn 
\EEA
Clearly, neglecting the difference between
triplices with different $\ell$ is an uncontrolled step which can (and
will) be improved in future work.

Given a set of integral equations \Ref.oneF2  , \Ref.oneTz  ~it is
customary to expand the unknown functions in an appropriate complete
set of functions, and to project the resulting expanded form on each
function in the set separately. In this way one generates a countable
set of algebraic equations. The least automatic step in this procedure
is the choice of the complete set of functions. By choosing the low
order functions in the basis set to represent in some sense the
properties of the expected solutions, one can hope that a truncated
set may serve as well.  It is of course also very helpful if the used
set of functions is simple enough to allow as much analytic
integration as possible. To find a good compromise between these
requirements we suggest to use a set of functions:
\BE\label{set}
f_0(\O)={1\over \O+i\g}\,, \dots
\quad f_n(\O)={d^n\over d \g ^n }{1\over \O+i\g}\,,\dots 
\EE
which correspond in $t$-representation to the set of Laguerre functions:
$\exp(-\g t)\,, \dots t^n\exp(-\g t)\dots $ With these functions one
may compute analytically all the needed integrals and they have a
reasonable asymptotic behavior for large $t$. 
  
The explicit solution of Eqs. \Ref.oneF2 , \Ref.oneTz ~is not a trivial
task and considerable amount of work will be called for in the future
to find and to analyze it. In this paper we proceed on qualitative
grounds to demonstrate that the solvability condition of these
equations contains the phenomenon of anomalous scaling. For this
purposes we will employ just the first non-vanishing term in the
set~\Ref.set ~ as displayed below:
\BEA
\label{trF} \tilde \C.F_{2;h}(\O)&\Rightarrow  &F{\g\over
  \O^2+\g^2}\,,
\\ \label{trip1}
\C.T_{A;h}^{(0)}(\Om,\Op)& \Rightarrow     &T{1\over
  (\Op-i\g_1)^2(\Om+i\g_1)^2}\ .
\EEA
One sees that for $\tilde \C.F_{2;h}(\O)$ we took Im$f_0(\O)$ as
required by the evenness of a real function $\tilde \C.F_{2;h}(\O)$.
In $t$-representation the triplex $\C.T_{A;h}(t_-,t_+)$ must be zero
if $t_+$ and/or $t_-$ are less or equal to zero. The homogeneous
evolution equations for the triplices dictate that they are small for
small $t_+$ and/or $t_-$.  Therefore there is no contribution of $f_0(\O)$
to the triplices, and we employ the representation~\Ref.trip1 ~ in which
the choice between $f_1(\O)$ and $f_1^*(\O)$ is dictated by 
causality.  We introduced in Eq.~\Ref.trip1 ~an additional freedom
($\g_1\ne\g$) which will allow to see later how qualitative results
depend on the choice of the trial functions. The results are
only weakly sensitive to the ratio $\g_1/\g$.

As one knows for example from quantum mechanical calculations of
energy levels, eigenvalues are usually much less sensitive to the choice 
of trial functions than the corresponding eigenfunctions. This provides
an additional justification to the expectation that our eigenvalue 
problem of finding
$\C.Z(h)$ from the solvability condition of Eqs.~\Ref.oneF2 , \Ref.oneTz
~is still meaningful even with very simple trial functions~\Ref.trF ,
\Ref.trip1 .  To formulate the solvability condition we consider a
consequence of Eq.~\Ref.oneTz ~which may be called their
``one-frequency'' reduction: multiply Eq.~\Ref.eqTz ~for $\tilde
\C.T_h^{(0)}$ by $\d (\O + \Om -\Op)$ and integrate over $\Op$ and
$\Om$.  This gives:
\BE\label{oneT} -[i \O +\tilde\sigma_h( \O )]\tilde
\C.T_{h}( \O)= \e^4 [\tilde \C.D_h(\O)+\tilde \C.E_h(\O)]\,, 
\EE
in which
\BEA\label{reduction}
\tilde \C.T_{h}( \O)&\equiv \int {d\Om d\Op\d (\O+ \Om
-\Op)\over 2\pi } \tilde \C.T_{h}^{(0)}
(\Om,\Op)\,,\\ \label{reductionD}
\tilde \C.D_{h}( \O)&\equiv \int {d\Om d\Op\d (\O+ \Om
-\Op)\over 2\pi }\tilde \C.D_{h}^{(0)}
(\Om,\Op)\,, \\      \label{reductionE}
\tilde \C.E_{h}( \O)&\equiv \int {d\Om d\O\d (\O+ \Om
-\Op)p\over 2\pi } \tilde \C.E_{h}^{(0)}
(\Om,\Op)\ .
\EEA

\begin{figure}\hskip -.3cm 
  \epsfxsize=18 truecm \epsfbox{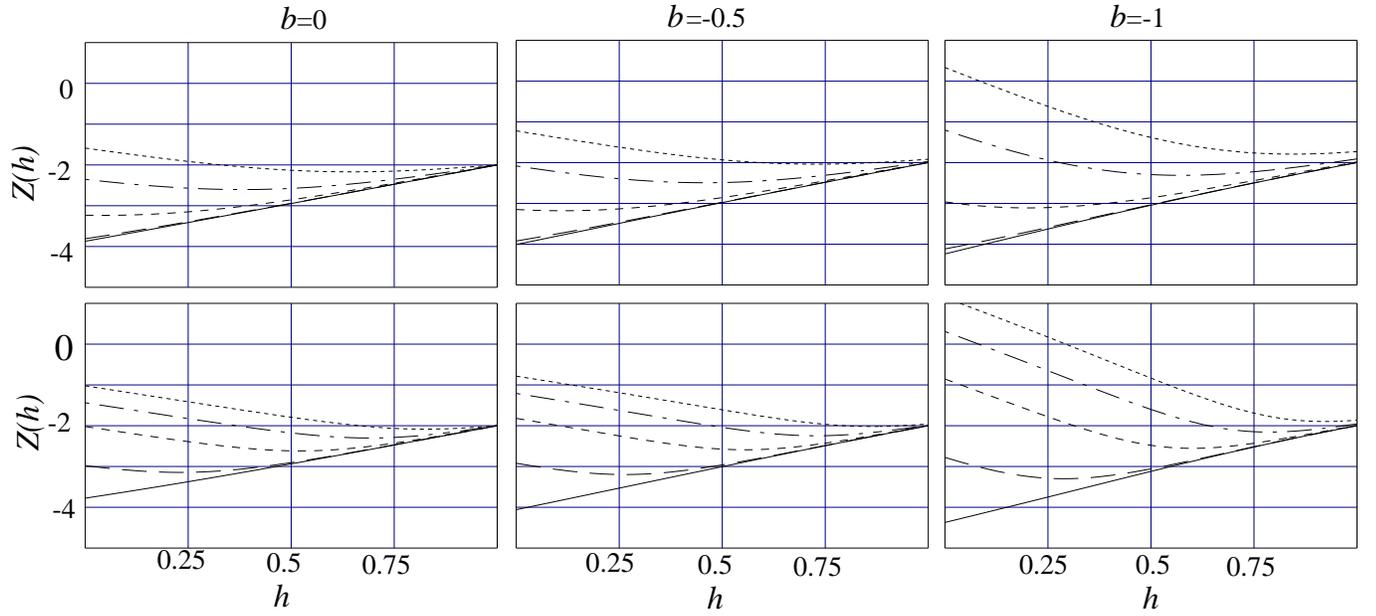} \vskip .2cm
\caption{The function $\C.Z(h)$ for different values of $\e$, $b$ and $\l$.
Different line types  denote different values of $\e$:
$\e=$0, 0.25, 0.5, 0.75 and 1 are shown with solid, long  dashed, dashed, 
dash-dotted and dotted lines respectively.  The three columns correspond to
$b=0,\, -0.5$ and $-1$, as shown over the panels. The upper panels correspond
to $\l=2$, the lower ones to $ \l=4$. The ratio $\gamma_1/\gamma$=0.5 . }
\label{f:Zh} 
\end{figure}

\begin{figure}\hskip -.3cm 
\epsfxsize=18 truecm
\epsfbox{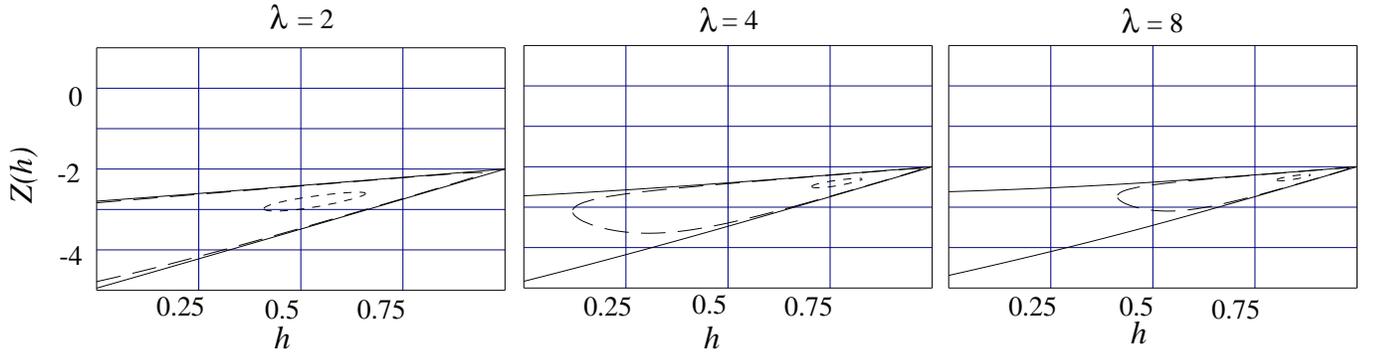}
\vskip .2cm 
\caption{The functions    $\C.Z(h)$  at $b=-2$ and  $\gamma_1/\gamma$=0.5. 
The line types correspond to 
those in the previous figure.  The values of $\l$ are shown over
the different panels.}
\label{f:Zh1} 
\end{figure}

Equation~\Ref.oneT ~has exactly the same factor $[i \O
+\tilde\sigma_h( \O )]$ in its LHS as Eq.~\Ref.oneF2 ~for $\tilde
\C.F_{2;h}( \O)$:
\BE
\label{oneF2a} 
-[i \O +\tilde\sigma_h( \O )]\tilde \C.F_{2;h}( \O)= \tilde \C.B_h(\O )\ 
.  \EE 
This allows the elimination of the non-homogeneous terms $\propto
\O$, multiplying Eq.~\Ref.oneF2a ~by $\C.T_{h}( \O)$ and Eq.~\Ref.oneT
~by $ \tilde \C.F_{2;h}( \O)$ and taking their difference:
\BE\label{solv1}
\tilde \C.T_{h}( \O)\tilde \C.B_{h}( \O)=\e^4 \tilde \C.F_{2;h}( \O ) 
 [\tilde \C.D_{h}( \O)
+\tilde \C.E_{h}( \O)]\ .
\EE
Recall that according to~\Ref.B ~$\tilde \C.B_{h}$ is a form linear in
the triplices $\C.T_{A;h}$ and quadratic in $\C.F_{2;h}$, while
$\tilde \C.D_{h}$ and $\tilde \C.D_{h}$ are forms quadratic in $\tilde
\C.T_{A;h}$ and linear in $\tilde \C.F_{2;h}$. Therefore
Eq.~\Ref.solv1 ~is a homogeneous form which is quadratic both in
$\tilde \C.T_{A;h}$ and $\tilde \C.F_{2;h}$.  With our
representation~\Ref.trF , \Ref.trip1 ~of $\C.F_{2;h}$ and $\tilde
\C.T_{A;h}$ Eq.~\Ref.solv1 ~is proportional to $F^2\,T^2$ which may be
canceled if there exists  a non-trivial solution of the problem.  
Therefore one
may think that Eq.~\Ref.solv1 ~is the solvability condition which we are
looking for. However, the trial functions~\Ref.trF ~and \Ref.trip1
~are not a solution of Eqs.~\Ref.oneF2a ,~\Ref.oneT . Therefore we
cannot expect to satisfy Eq. \Ref.solv1 ~for all values of the
frequency. According to our calculational scheme we have to project
the initial Eqs.~\Ref.oneT , \Ref.oneF2a ~ on the (truncated) set
of functions $f_0(\O)$,
$f_1(\O)$ and consider the algebraic consequences of this step.  Actually
Eq.~\Ref.solv1 ~sheds light on how we may proceed in order to find a
solvability condition in our scheme.  Namely, we may project
Eq.~\Ref.oneF2a ~for $\C.F_{2;h} $ on the triplex trial
function~\Ref.trip1 ~and Eq.~\Ref.oneT ~for $\tilde \C.T_{A;h}$ on the
trial function~\Ref.trF ~for $\C.F_{2;h} $ and subtract the
results.  This procedure corresponds to $\O$-integration of
Eq.~\Ref.solv1 :
\BEA
\label{solv2} &&
\int{d\O\over 2\pi} \tilde \C.T_{h}( \O)\tilde \C.B_{h}(
  \O)\\
&=&\e^4 \int{d\O\over 2\pi}  [\tilde \C.F_{2;h}( \O ) \tilde \C.D_{h}( \O)
+\tilde \C.F_{2;h}( \O ) \tilde \C.E_{h}( \O)]\ .\nn
\EEA
In Eq.~\Ref.solv2 ~$\tilde \C.B_{h}( \O)$ is given by Eqs.~\Ref.B ~--
\Ref.Vc , $\tilde \C.D_{h}( \O)$ and $\tilde \C.E_{h}( \O)$ are
defined by Eqs.~\Ref.reductionD ~and \Ref.reductionD ~in which $\tilde
\C.D_{h}^{(0)}( \Om,\Op) $ and $\tilde \C.E_{h}^{(0)}( \Om,\Op) $
are given by \Ref.ED ~-- \Ref.Wc . All these equations contain the functions
$\tilde \C.F_{2;h}( \O ) $ and $\tilde \C.T_{h}( \O)$ for which we
will use the representations \Ref.trF ~and \Ref.trip1 . Finally
Eq.~\Ref.solv2  ~yields:
\BE\label{solv3}
{a\over  R_{_\C.Z}} U_{a}(h) +b\,U_{b}(h)+
cR_{_\C.Z} U_{c}(h)\,=0\,,
\EE
where $R_{_\C.Z}=\l ^{2+\C.Z(h)}$ and $U_{a}(h)$, $U_{b}(h) $ and
$U_{c}(h) $
are $\O$-integrals of $U_{a;h}(\O)$, $U_{b;h}(\O) $ and $U_{c;h}(\O)$ and
\BEA
U_{a;h}(\O)&  \equiv  &V_{a;h}(\O)-\e^4 W_{a;h}(\O)\,, \
U_{b;h}(\O)\equiv V_{b;h}(\O)\,, \nl            \label{solv5}
U_{c;h}(\O)&\equiv     &V_{c;h}(\O)-\e^4 W_{c;h}(\O)\ .
\EEA
Equation~\Ref.solv3 ~are quadratic with respect to $ R_{_\C.Z}=\l
^{2+\C.Z(h)}$ and allows to find $\C.Z$ as a function of $h$.  The actual
calculation has been done with the help of {\em Matematica}, a system
for doing analytical calculations by a computer\cite{wol}. The resulting set
of functions $\C.Z(h)$ for $\e=0,\,0.25,\,0.5,\,0.75,\,1$, for the spacing parameters
$\l=2,\,4$ and the parameter $b=0,\,-0.5,\,-1$ is shown  in
Fig.~\ref{f:Zh}. Figure~\ref{f:Zh1} displays the functions $\C.Z(h)$ for
the same set of $\e$, $b=-2$ and $\l=2,\,4,\,8$.
Note that at for $\epsilon=0$ (solid lines) in both Figs.10 and 11 at $h=1$ all the
functions $\C.Z(h)$ are equal to $-2$ (this is relatively easy to see 
analytically).  For $h<1$ one sees that the functions $\C.Z(h)$ go down
almost as straight lines.  An important conclusion is that at $\e=0$
the functions $\C.Z(h)$ are always negative (moreover, $\C.Z(h)\le -2$).
Recall that the scaling exponents $\z_n$ are related with $\C.Z(h)$
via saddle-point requirement~\Ref.saddle ~ and, in particular,
$\z_0=\min_h \C.Z(h)$. By definition, the zero-order structure function is
1, therefore $\z_0=0$ and, consequently, for physical solutions $\C.Z(h)\ge 0$.
Negative values of $\C.Z(h)$ are unphysical like
the negative values of masses which may appear as a solution for
equilibrium of a mechanical system. Remember that in
the derivation of the closure equations we have neglected
inhomogeneous terms with normal K41 scaling with respect of terms
with anomalous scaling; this is only possible if $\C.Z(h)>0$.
Therefore for negative values of $\C.Z(h)$ the ``anomalous" solutions must be
disregarded in favor of solutions with
normal scaling. The  first important conclusion is therefore that {\em at  
$\e=0$ the qualitative analysis of the controlled closure predicts 
K41 scaling}.

We next consider the parameter range for which $a\,c<0$. This 
mimics the Navier Stokes case in directing the cascade from
large to small scales. The parameter $b$ is chosen in the  region 
of stability $-2 < b<0$, for
which the effect of the flux of the second (``helicity") integral of motion becomes
irrelevant deep in the inertial interval. With $a$ taken to be unity,
these two constraints limit $b$ to the interval $[-1,0]$. Indeed, in
most previous simulations the value $b=-0.5$ was chosen.
The calculated functions $\C.Z(h)$ for this region of $b$ are exhibited in
Fig.~\ref{f:Zh}. One sees that in all the panels the values of $\C.Z(h)$
increase with $\e$ and some of the functions
$\C.Z(h)$ become positive in the left part of the $h$ interval.
Moreover, $\C.Z(h)$ gains a minimum (although still in the negative
region). All these qualitative features are precisely those expected
for the functions $\C.Z(h)$ in the case of
anomalous scaling. We note that the present level of analysis
cannot yield quantitative results. The functions  $\C.Z(h)$ are
still mostly negative, and we only see the good trend of creating
a minimum and inching up toward positive territory with the 
increase in $\epsilon$.  Although we cannot discuss at  present
the actual values of the anomalous scaling
exponents, their dependence on the parameters of the model: $\e$,
$\l$ and $b$, etc, we think that results in
the Fig.~\ref{f:Zh} can be considered as definite evidence for the birth 
of anomalous
scaling with the increase of the value of $\e$.  An additional
encouraging result is the finding that in the non-physical interval of $b$ (see in
Fig.~\ref{f:Zh1}) we do not see any of the good phenomena discussed
above: at $\e=0$ we have
two negative branches of $\C.Z(h)$ and with increasing of the value of 
$\e$ the solutions stay in between these branches without a tendency to
increase.  Moreover for $\e >1/2$ the solutions  $\C.Z(h)$ disappear altogether.

%%%%%%%%%%%%%%%%
\end{multicols}\widetext
%%%%%%%%%%%%%%%%%%%%%%%%%%%%%%%%
\begin{multicols}{2}

\section{Summary and Discussion}
For the sake of clarity we summarize verbally the important
steps in this paper.
\begin{enumerate}
\item We first introduced the $(N,\epsilon)$-shell model in which $N$
  copies of the ``Sabra" shell model are coupled with the interaction
  amplitudes shown in Eq.(\ref{choice}). These amplitudes contain a
  deterministic and a random part with the parameter $\epsilon$
  determining their relative importance.
\item For $\epsilon=0$ the model reduces to the well known random
  coupling model \cite{61Kra}, in which there is no anomalous scaling
  in the limit $N\to \infty$. For $\epsilon=1$ the system boils down
  to $N$ uncoupled copies of the ``Sabra" model with anomalous scaling
  for each.
\item We studied numerically the behavior of the model for values of
  $\epsilon$ in the interval $[0,1]$ in the limit $N\to \infty$.  The
  results are shown in Figs.1 and 2. For $\epsilon=0$ we find no
  anomalous scaling as expected, for $\epsilon=0.8$ we find anomalous
  corrections smaller than those found for the Sabra model which
  pertains to $\epsilon=1$.
\item We derived the linear hierarchy of evolution equations for the
  correlation and response functions of this model. In the inviscid
  limit the hierarchy of equations for the correlation functions is
  linear and homogeneous.
\item We pointed out that the homogeneity of the hierarchy of
  equations results in a rescaling symmetry of these equations which
  foliates the solutions on so-called $h$-slices, where $h$ is the
  rescaling exponents of the velocity variable. The $n$th order
  velocity correlation function rescales on an $h$-slice with a
  scaling exponent $nh+{\cal Z}(h)$, where ${\cal Z}(h)$ is associated
  with the rescaling properties of the probability measure.
\item The full solution is written as an integral over contributions
  with different $h$, and leads to multiscaling via the saddle point
  calculation of the corresponding integrals over $h$.  The values of
  the scaling exponents are determined by the function ${\cal Z}(h)$
  according to Eq.(\ref{saddle}).
\item The hierarchical equations written on an $h$ slice can be closed
  in a controlled fashion using $\epsilon$ as a small parameter. We
  derived the closed equations for the 2nd and 3rd order objects,
  retaining terms up to $O(\epsilon^4)$ and showing that the parts
  neglected are of $O(\epsilon^6)$. We could go a further step in the
  closure scheme by writing down a system for the 2nd, 3rd and 4th
  order objects, retaining terms up to $O(\epsilon^6)$ and neglecting
  terms of $O(\epsilon^8)$ etc.
\item Section \ref{s:answer} presents a preliminary analysis of ${\cal
    Z}(h)$ from solvability condition for the closed set of equations
  for the 2nd and 3rd order objects. The full analysis is technically
  involved and calls for further work.  The main qualitative
  conclusions of the present analysis is that there exists no
  anomalous solution for $\epsilon=0$, whereas for larger values of
  $\epsilon$ we see the birth of anomalous scaling. The level of
  precision is currently not sufficient for quantitative comparisons
  with the numerical simulations, but we judge the results extremely
  encouraging for further work in the near future.
\end{enumerate}
In summary, we presented above what appears to be the first
attempt to compute from the equations of motion the anomalous exponents 
of the correlations of a nonlinear turbulent system in a controlled fashion. 
The small parameter $\epsilon$ has been used to close the infinite hierarchy
of equations for the statistical objects, from which one can compute
the 2nd and 3rd correlation functions explcitly. Nevertheless we considered
the calculation of ${\cal Z}(h)$ only, and also this calculation turned
out to be too hard to be performed exactly. The bottom line of this
analysis is that (i) the anomaly of the scaling exponents is analytically
predictable by the method developed above, and (ii) the actual calculation
of numbers is rather difficult and calls for additional careful work.

\acknowledgments We thank Barak Galanti for his assistance with the
numerical simulations, and to Gregory Eyink for sharing with us his
unpublished results in \cite{98Eyi}. The work has been supported in
part by the European Commission under the Training and Mobility of
Researchers program, The German-Israeli Foundation, the Israel Science
Foundation administered by the Israel Academy of Sciences, and the
Naftali and Anna Backenroth-Bronicki Fund for Research in Chaos and
Complexity.
\end{multicols}
\appendix
\section{Sketch of Derivation of the hierarchy of evolution equations}
\label{a:hier}%  \vskip -.5cm \rightline{\sf a:hier }\vskip   .5 cm 
\subsection{Hierarchy for the correlation functions}
\label{aa:hierF}
%\vskip -.5cm \rightline{\sf aa:hierF}\vskip   .5 cm 
Let us compute $\partial F_{2;\s,\s}^\a(k_n;t-t')/\partial t$ defined
by Eq.~\Ref.F2 ~at $t'=0$, substituting instead of $d u_{n,\s}^\a(t)/
d t$ the RHS of Eq.~\Ref.sabrafourier . With the help of
definition~\Ref.F3 ~the result may be written as: \BEA &&
\frac{\partial }{\partial t}F_{2;\s ,\s}^{~~\a}(k_n;t ) =\sum_{\b,\g
  ,\s',\s''} \Phi^{\g,\b,\a} [\D_{\a+\b,\g}+ \D_{\a+\b+N,\g}+
\D_{\a+\b,\g+N} ]A^{\s}_{\s'\s''} \nl &\times& \big \{ ak_{n+1}
F_{3,\s\s'\s''}^{\a,\b,\g}(k_{n+1};0,t,t)+ b k_{n}
F_{3,\s\s'\s''}^{\a,\b,\g}(k_{n};t,0,t) + ck_{n-1}
F_{3,\s\s'\s''}^{\a,\b,\g}(k_{n-1};t,t,0)\big\}\,,
\label{eq:hier1}
\EEA where we used the symmetry properties $
A^{\s}_{\s'\s''}=A^{\s'}_{\s\s''}$.   Summing this equation over $\a$
and $\s$ we derive Eqs.~\Ref.hierF3 ~with the invariant
combination~\Ref.F3a ~ in the RHS.
\subsection{Definitions of the Green's function}
\label{aa:defG}\vskip -.5cm \rightline{\sf aa:defG}\vskip   .5 cm 
First we define the ``tensorial'' Green's function
\BE
G_{1,1;\s,\s'}^{\a,\b} (k_n;t_1-t_2)=
\left\langle
\delta u_{n,\sigma}^\alpha(t_1) \over
\delta f_{n,\sigma'}^{\b*}(t_2)
\right\rangle \, , \label{G11t}
\EE
and its scalar counterpart which is invariant under all the symmetry
transformations described in Subsect.~\ref{ss:symmetry}:
\BE
G_{1,1}(k_n;t_1-t_2)\equiv  {1\over N}\sum_{\a,\s}
G_{1,1;\s,\s}^{\a,\a} (k_n;t_1-t_2)\ .
\EE
Next we define the 3rd order tensorial Green's functions $\B.G_{2,1}$
and $\B.G_{1,2}$ describing a response of two velocities on one
forcing and {\em vice versa}:
\BEA \label{G21t}
G_{2,1;\s\s'\s''}^{\a,\b,\g}(k_{n},k_{m},k_{\ell};t_n;t_m,t_\ell)&\equiv&
\left\langle \frac{\d [u_{m,\s'}^{\b}(t_m) u_{l,\s''}^{\g }(t_\ell)]}
  {\d f_{n,\s}^{\a *}(t_n)} \right\rangle \,,\\
G_{1,2;\s\s'\s''}^{\a,\b,\g}(k_{n},k_{m},k_{\ell};t_n,t_m,t_\ell)&\equiv&
\left\langle \frac{\d ^2  u_{l,\s''}^{\g }(t_\ell)}
  {[\d f_{m,\s'}^{\b}(t_m) \d f_{n,\s}^{\a }(t_n)]^*} \right\rangle \ .\label{G12t}
\EEA
Analogously to the definition of the scalar 3rd order correlation
function~\Ref.F3a ~we define here the scalar counterparts of these 3rd order
Green's functions:
\BEA
 G_{2,1}(k_{n},k_{m},k_{\ell};t_n;t_m,t_\ell)&\equiv&
{1\over N} \sum_{a, \beta,\g} \sum_{\s,\s'\s''}
\Phi^{\g,\b,\a} A^{\s}_{\sigma'\sigma''}[\D_{\a,\b+\g} +\D_{\a+N,\b+\g}  +
\D_{\a,\b+\g+N}]\nl
&&\times
G_{2,1;\s\s'\s''}^{\a,\b,\g}(k_{n},k_{m},k_{\ell};t_n;t_m,t_\ell)\,,
\label{G21s}\\
G_{1,2}(k_{n},k_{m},k_{\ell};t_n;t_m,t_\ell)&\equiv& {1\over N}
\sum_{a, \beta,\g} \sum_{\s,\s'\s''} \Phi^{\g,\b,\a}
A^{\s}_{\sigma'\sigma''}[\D_{\a,\b+\g} +\D_{\a+N,\b+\g} +
\D_{\a,\b+\g+N}]\nl &&\times
G_{1,2;\s\s'\s''}^{\a,\b,\g}(k_{n},k_{m},k_{\ell};t_n,t_m,t_\ell)\ 
.\label{G12s} 
\EEA
As we discussed in Subsect.~\ref{ss:def-triplex} any three point
objects of the Sabra-shell model differ from zero only for three
consecutive shell numbers. This allows one to consider instead of each
Green's function \Ref.G12s ~and \Ref.G21s ~of three $k$ arguments three
functions with just one (middle) $k$ argument. Exactly like in the
case of triplices we introduce superscripts $^{(0)}$ and $^{(\pm)}$ to
distinguish them.  Using the convention discussed in
Subsect.~\ref{ss:def-triplex} we can define $G_{2,1}^{(0)}$,
$G_{2,1}^{(\pm)}$,  $G_{2,1}^{(0)}$ and $G_{2,1}^{(\pm)}$. For example:
\BE
G_{2,1}^{(0)}(k_{m};t_{m-1};t_m,t_{m+1})\equiv
G_{2,1}(k_{m},k_{m-1},k_{m+1};t_m;t_{m-1},t_{m+1}) \ .
\EE
%%%
Similarly we define the successive scalar response functions of the 4th order
as follows:
\begin{eqnarray}
&&G_{3,1}(k_{n1},k_{n2},k_{n3},k_{n4};t_1,t_2,t_3,t_4)
\equiv
{1\over N^2}
A^{\sigma'}_{\sigma''\sigma_3}A^{\sigma'}_{\sigma_4\sigma_5}
\left\langle {
\delta ( u_{n1,\sigma''}^{\beta *}(t_2)
u_{n2,\sigma_3}^{\alpha}(t_3)
u_{n3,\sigma_4}^{\alpha *}(t_4))
\over
\delta f_{n1,\sigma'}^{\beta}(t_1)
}\right\rangle
\, ,\label{G31}\\
&&G_{2,2}(k_{n1},k_{n2},k_{n3},k_{n4};t_1,t_2,t_3,t_4)
\equiv
{1\over N^2}
A^{\sigma'}_{\sigma''\sigma_3}A^{\sigma'}_{\sigma_4\sigma_5}
\left\langle {
\delta ( u_{n2,\sigma''}^{\beta *}(t_2)
u_{n3,\sigma_3}^{\alpha}(t_3))
\over
\delta f_{n1,\sigma'}^{\beta}(t_1)
\delta f_{n4,\sigma_4}^{\alpha *}(t_4)
}\right\rangle
\, ,\cdots \label{G22}
\end{eqnarray}
with the convention of sums over repeated indices.  Note that all the
$G$'s are real functions.  Again we note that it is more convenient to
write the hierarchical equations in the terms of quantities summed
over $\sigma$'s. These are invariant under all the symmetry
transformations discussed in Sect.~\ref{ss:symmetry}.

\subsection{Hierarchical equations for the Green's  functions}
\label{a:eq-Green's}%\vskip -.5cm \rightline{\sf  a:eq-Green's }\vskip   .5 cm 
The tensorial equation for the 2nd order $G_{1,1;\s\s}^{\a,\a}$ reads:
\BEA
\label{eq:hier2}
 \frac{\partial }{\partial t}G_{1,1;\s
  ,\s}^{\a,\a}(k_n;t ) &=&\sum_{\b,\g ,\s',\s''}
\Phi^{\g,\b,\a} [\D_{\a+\b,\g}+   \D_{\a+\b+N,\g}+    \D_{\a+\b,\g+N}
]\big \{ A^{\s}_{\s'\s''} \big[
\g_{a,n+1}
G_{2,1;\s\s'\s''}^{\a,\b,\g }(k_n,k_{n+1},k_{n+2};0,t,t)  \nl
&& + \g_{b,n}
G_{2,1;\s\s'\s''}^{\a,\b,\g }(k_{n},k_{n-1},k_{n+1};0,t,t)\big] + C^{\s}_{\s'\s''}
\g_{c,n-1} G_{2,1;\s\s'\s''}^{\a',\b,\g}(k_{n},k_{n-2},k_{n-1}
;0,t,t)\big\}+ \d(t)\ .
\EEA
By following more or less the same procedure as for the correlation
functions in Appendix~\ref{aa:hierF} we get the hierarchical equations:
\begin{equation}
{\partial \over \partial t}
G_{1,1}(k_n;t)
= \gamma_{a,n+1}
G_{2,1}^{(-1)}(k_{n+1};0,t,t)
\nonumber \\
\gamma_{b,n}
G_{2,1}^{(0)}(k_{n};t,0,t)
+ \gamma_{c,n-1}
G_{2,1}^{(+1)}(k_{n-1};t,t,0)+ \d(t)\ .
\label{hierG11} 
\end{equation}
Analogously for the higher order Green's functions one has:
 \begin{eqnarray}
{\partial \over \partial t_1}
G_{2,1}^{(0)}(k_n;t_1,t_2,t_3)&= &\gamma_{a,n}
G_{3,1}(k_n,k_n,k_{n+1}k_{n+1};t_2,t_1,t_1,t_3)+ 
\gamma_{b,n-1}
G_{3,1}(k_n,k_{n-2},k_nk_{n+1};t_2,t_1,t_1,t_3)\nl
&&+\gamma_{c,n-2}
G_{3,1}(k_{n},k_{n-3},k_{n-2}k_{n+1};t_2,t_1,t_1,t_3)\,,\\
{\partial \over \partial t_1}
G_{1,2}^{(-1)}(k_n;t_1,t_2,t_3)&=&
\gamma_{a,n  }
G_{2,2}(k_n,k_{n+1},k_n, k_{n+1};t_1,t_1,t_2,t_3)+
\gamma_{b,n-1}
G_{2,2}(k_{n-2},k_n,k_nk_{n+1};t_1,t_1,t_2,t_3)\nl  &&+
\gamma_{c,n-2}
G_{2,2}(k_{n-3},k_{n-2},k_n,k_{n+1};t_1,t_1,t_2,t_3) \,, 
\end{eqnarray}
 and similarly for the other $G_{2,1}$'s and $G_{1,2}$'s.\\
%%%%%%%%%%%%%%%%%%%%%%%%%%
\begin{multicols}{2}
%%%%%%%%%%%%%%%%%%%%%%%%%%%%%%%%%%%%
\section{The closure equations in the $A$-triplex approximation}
\label{a:Aclos}%\vskip -.5cm \rightline{\sf  a:Aclos }\vskip   .5 cm 
\subsection{Equations in the $\o$-representation}
\label{aa:Aomega}%\vskip -.5cm \rightline{\sf  aa:omega }\vskip   .5 cm 

Given a stationary case, when all the functions depend on time
differences only, it is useful to consider the closure equations in
the $\o$-representation. Define:
\BEA\label{omega-F} \tilde
F_{2;h}(k_n;\o) &=& \int\limits_{-\infty}^\infty
F_{2;h}(k_n;t)\exp (i\o t )dt \,, \\
\tilde F_{2;h}^*(k_n;\o) &=& \int\limits_{-\infty}^\infty F_{2;h}^*(k_n;t)\exp
( - i\o t )dt\ . \\ \nn 
\EEA 
As in these equations, we will use tilde signs over a
character to denote a Fourier transform of that function. It follows from
Eq.~(\ref{omega-F}) that
 \BE\label{backF}
F_2(k_n;t) = \int\limits_{-\infty}^\infty \tilde F_2(k_n;\o )\exp (-
i\o t ){d \o \over 2\pi}\ .  
\EE
According to the definition~(\ref{F2},\ref{F2a})
$F_{2;h}(k_n;t)=F_{2;h}^*(k_n;-t)$.  Therefore {\sl $\tilde
  F_{2;h}(k_n;\o) $ is a real function of $\o $}.

Next we define the Fourier transform $\tilde T_{h}$ of any 3rd order
object $T_{h}$ as follows: \BEA \nn &2&\pi \tilde T_{h}
(k_n;\om ,\oz ,\op )\delta (\om + \oz - \op )
=\int\limits_{-\infty}^{\infty} dt_-dt_0 dt_+ \\
&&\times T_{h} (k_n;t_-,t_0,t_+)\exp[i(\om t_- +\oz t_0 -\op t_+)] \ .
\label{omega-F3} \EEA Recall that with our convention (see
Subsect.~\ref{ss:def-triplex}) $t_0$ and $t_\pm$ are times of the $n$-
and $(n \pm 1)$-shells.  With this definition the sum of incoming
frequencies $\om + \oz$ is equal to the outgoing frequency $\op $. 

The equations for the 2nd order correlation function (two top lines in
Fig.~\ref{f:closed1}) and for triplices (two last lines) in the
$\o$-representation may be written as: 
\end{multicols}
%%%%%%%%%%%%%%
\leftline{-----------------------------------------------------------------------------}
\BEA
\label{eqf2} 
-[i \o +\tilde   \Sigma_h(k_n;\o )]\tilde F_{2;h}(k_n;\o)&=& \tilde
B_h(k_n,\o )\,, \\
\label{omega-Tp}
[i\op +\tilde \Sigma_h^*(k_n;\o _+) ] \tilde T_h^{(+1)}
(k_{n-1};\o _-,\o_ 0)&=&\e^4 \big[
 \tilde D_h^{(+1)}(k_{n-1};\o_ -,\o_ 0 )  + \tilde E_h^{(+1)}(k_{n-1};\o_ -
,\o_ 0 ) \big] \,, \\
-[i \oz +\tilde \Sigma_h(k_n;\oz )]\tilde
T_{A;h}^{(0)}(k_n;\om,\op)&=& \e^4 \big[ \tilde E_h^{(0)}(k_n,\om,\op )
+ \tilde D_h^{(0)}(k_n,\om,\op )\big] \,, 
  \label{omega-Tz}\\
-[i\om +\tilde  \Sigma_h(k_n;\o _- )]
 \tilde  T_{A;h}^{(-1)}(k_{n+1};\o_0,\o_+)&=&\e^4 \big[
\tilde  D_h^{(-1)}(k_{n+1};\o_0,\o_+ ) +  \tilde
  E_h^{(-1)}(k_{n+1};\o_0,\o_+ )\big]\ . \label{omega-Tm}
\EEA
Note that the parameter $\epsilon$ appears explicitly here, as it was
discussed in Sect.\ref{ss:mag}.
Here the ``mass-operator'' $ \tilde \Sigma_h(k_n;\o ) $ originates from
the diagram A in Fig.~\ref{f:closed1} and may be written as:
 \BEA
\label{omega-s} && \tilde \Sigma_h(k_n;\o)= {1\over
  2\pi}\int\limits_{-\infty}^\infty
d\om d\oz d\op \delta({\om + \oz - \op}) \\
&& \times \Big\{ \gamma_{n+1,-1}\Big[\tilde F_{2;h}(k_{n+2};\op ) \tilde
T_{A;h}^{(0)*}(k_{n+1}; \om , \op ) + \tilde F_{2;h}(k_{n+1};\oz ) \tilde
T_{A;h}^{(+1)*}(k_{n+1};
\om , \oz )\Big]\delta(\o - \om) \nn \\
&&+ \gamma_{n,0}\Big[\tilde F_{2;h}(k_{n+1};\op ) \tilde T_{A;h}^{(-1)*}
(k_{n}; \oz , \op )   + \tilde F_{2;h}(k_{n-1};\om ) \tilde
T_{A;h}^{(+1)*}(k_{n};
\om , \oz )\Big]\delta(\o - \oz) \nn \\
&&+ \gamma^*_{n-1,1}\Big[\tilde F_{2;h}(k_{n-1};\oz ) \tilde T_{A;h}^{(-1)}
(k_{n-1}; \oz , \op )   + \tilde F_{2;h}(k_{n-2};\om ) \tilde
T_{A;h}^{(0)}(k_{n-1}; \om , \op ) \Big]\delta(\o - \op) \Big\} \ . \nn
\EEA 
Here and below we omit, for the sake of brevity, the frequency argument of the
leg with the Green's function in $A$-triplices  and the corresponding argument in
$D$ and $E$ functions. It may be found from the relationship
$\op=\om+\oz$.   The term $\tilde B_h(k_n,\o )$ in the RHS of
Eq.~\Ref.eqf2 ~corresponds to the diagram B in Fig.~\ref{f:closed1}:
\BEA
\label{omega-B} \tilde B (k_n;\o)&=&
{1\over 2\pi}\int\limits_{-\infty}^\infty
d\om d\oz d\op 
\Big\{ \gamma_{n+1,-1} \tilde F_{2;h}(k_{n+1};\oz ) \tilde
F_{2;h}(k_{n+2};\op ) \tilde T_{A;h}^{(-1)*}(k_{n+1};
\oz , \op )\delta(\o - \om) \\
&&+ \gamma_{n,0} \tilde F_{2;h}(k_{n-1};\om ) \tilde F_{2;h}(k_{n+1};\op )
\tilde T_{A;h}^{(0)*}(k_{n}; \om , \op )  \delta(\o - \oz) \nl
&&+ \gamma_{n-1,1}^* \tilde F_{2;h}(k_{n-2};\om ) \tilde F_{2;h}(k_{n-1};\oz )
\tilde T_{A;h}^{(+1)}(k_{n-1}; \om , \oz )\delta(\o - \op )\Big\} \delta({\om
  + \oz - \op}) \ .
   \nn 
\EEA
Functions $D$ correspond to the diagram D:
\BEA
\label{DDp}
 \tilde D_h^{(+1)} (k_{n-1};\om, \oz )&=&
\int\limits_{-\infty}^\infty {d\o \over 2\pi}
\Big\{  \gamma_{n,0}^* \tilde T_{A;h}^{(+1)}(k_{n}; \oz , \o  )
 \tilde T_{A;h}^{(0)}(k_{n-1}; \om , \o  )
\tilde F_{2;h}(k_{n};\o  )   \\ \nn
&&+ \gamma_{n-1,1} \tilde T_{A;h}^{(0)*}(k_{n-2};\o ,\oz  )
 \tilde T^{+1}_{A;h}(k_{n-2};\o  , \om )
\tilde F_{2;h}(k_{n-3};\o  ) \Big\}\,,\\
\label{DDz}
 \tilde D_{h}^{(0)}(k_{n};\om, \op )&=&
\int\limits_{-\infty}^\infty {d\o \over 2\pi}
\Big\{  \gamma_{n+1,-1} \tilde T^{+1*}_{A;h} (k_{n+1};\o , \op )
 \tilde T_{1,h}(k_{n}; \om , \o  )
\tilde F_{2;h}(k_{n};\o  )   \\ \nn
&&+ \gamma_{n-1,1}^* \tilde T_{A;h}^{(+1)}(k_{n};\o ,\op  )
 \tilde T_{A;h}^{(+1)}(k_{n-1};\om ,\o   )
\tilde F_{2;h}(k_{n};\o  )  \Big\}\,, \\
\label{DDn}
 \tilde D_h^{(-1)} (k_{n+1};\oz, \op )&=&
\int\limits_{-\infty}^\infty {d\o \over 2\pi}
\Big\{  \gamma_{n,0} \tilde T_{A;h}^{(0)}(k_{n+1};\o ,\op  )
 \tilde T_{A;h}^{(-1)*}(k_{n};\o ,\oz  )
\tilde F_{2;h}(k_{n};\o  )  \\ \nn
&&+ \gamma_{n+1,-1} \tilde T_{A;h}^{(-1)*}(k_{n+2};\op , \o  )
 \tilde T_{A;h}^{(0)}(k_{n+2};\oz , \o  )
\tilde F_{2;h}(k_{n+3};\o   )  \Big\}\ .
\EEA
Functions $E$ correspond to the diagrams E$\a$ and E$\b$ in
Fig.~\ref{f:closed1}:
 \BEA \label{EEp}
 \tilde E_h^{(+1)} (k_{n-1};\om, \oz )&=&
\int\limits_{-\infty}^\infty {d\o \over 2\pi}
\Big\{  \gamma_{n,0}^*\tilde F_{2;h}(k_{n+1};\o  )
\tilde T_{A;h}^{(0)}(k_{n-1};\om,  \o -\oz  )
 \tilde T_{A;h}^{(0)}(k_{n}; \oz , \o  ) \\ \nn
&&+\gamma_{n-1,1}\tilde F_{2;h}(k_{n-2};\o  )
\tilde T_{A;h}^{(+1)}(k_{n-2};\oz- \o ,\om )
 \tilde T_{A;h}^{(-1)*}(k_{n-2};\o , \oz  ) \nl
&& + \gamma_{n,0}^*\tilde F_{2;h}(k_{n-1};\o  )
\tilde T_{A;h}^{(+1)}(k_{n};\oz,  \o +\om  )
 \tilde T_{A;h}^{(+1)}(k_{n-1}; \om , \o  )  \\ \nn
&&+\gamma_{n-1,1}\tilde F_{2;h}(k_{n-1};\o  )
\tilde T_{A;h}^{(0)*}(k_{n-2};\o  - \om ,\oz )
 \tilde T_{A;h}^{(+1)}(k_{n-2};\om , \o   )  \Big \} \,, \nn  \\
\label{EEz}
 \tilde E_0 (k_{n};\om, \op )&=&
\int\limits_{-\infty}^\infty {d\o \over 2\pi}
\Big\{  \gamma_{n+1,-1}\tilde F_{2;h}(k_{n+2};\o  )
\tilde T_{A;h}^{(+1)}(k_{n};\om,  \o -\op  )
 \tilde T_{A;h}^{(-1)*}(k_{n+1}; \op , \o  )  \\ \nn
&&+\gamma^*_{n-1,1}\tilde F_{2;h}(k_{n-1};\o  )
\tilde T_{A;h}^{(+1)}(k_{n-1};\om,\op - \o  )
 \tilde T_{A;h}^{(0)}(k_{n};\o , \op  ) \\ \nn
&&+   \gamma_{n+1,-1}\tilde F_{2;h}(k_{n+1};\o  )
 \tilde T_{A;h}^{(+1)*}(k_{n+1};\o  - \om, \op  )
 \tilde T_{A;h}^{(0)}(k_{n}; \om , \o  )   \\ \nn
&&+\gamma^*_{n-1,1}\tilde F_{2;h}(k_{n-2};\o  )
\tilde T_{A;h}^{(+1)}(k_{n};\o  + \om,\op )
 \tilde T_{A;h}^{(+1)}(k_{n-1};\o ,\om   ) \Big \} \,, \nn  \\
\label{EEn}
 \tilde E_{A;h}^{(+1)} (k_{n+1};\oz , \op )&=&
\int\limits_{-\infty}^\infty {d\o \over 2\pi}
\Big\{  \gamma_{n,0}\tilde F_{2;h}(k_{n+1};\o  )
\tilde T_{A;h}^{(-1)*}(k_{n};\op - \o , \oz  )
\tilde T_{A;h}^{(+1)}(k_{n+1}; \o , \op )  \\ \nn
&&+\gamma_{n+1,-1}\tilde F_{2;h}(k_{n+1};\o  )
\tilde T_{A;h}^{(0)}(k_{n+2};\oz,\o +\op )
 \tilde T_{A;h}^{(+1)*}(k_{n+2};\o  , \op  ) \\ \nn
&&+  \gamma_{n,0}\tilde F_{2;h}(k_{n-1};\o  )
\tilde T_{A;h}^{(0)}(k_{n+1};\oz - \o  ,\op  )
 \tilde T_{A;h}^{(0)*}(k_{n}; \o_4 , \oz ) \\ \nn
&&+\gamma_{n+1,-1}\tilde F_{2;h}(k_{n+2};\o  )
\tilde T_{A;h}^{(-1)*} (k_{n+2};\op ,\oz + \o  )
 \tilde T_{A;h}^{(+1)} (k_{n+2};\oz , \o   )  \Big \} \ . \nn
\EEA
%%%%%%%%%%%%%%%%%%%%%%%%%

\subsection{Closure equations for Scale-invariant Functions}
All the objects appearing in the previous equations
(\ref{eqf2})-(\ref{EEn}) are scale invariant, as they are defined in
an ``$h$-slice''.  We make use of this invariance to simplify the
closed set of equations by introducing the following scale invariant
dimensionless representation:
\BEA
\tilde F_{2;h}(k_n;\o )&\equiv &
{U \over k_0} \l ^{-(h+{\cal Z}(h)+1) }
\tilde \C.F_{2;h}\Big( \frac{\o}{\o_{n;h}}\Big)\,, \quad     
\tilde \S_h(k_n;\o)=\o_{n;h}\tilde \s_h  \Big( \frac{\o}{\o_{n;h}}\Big)\,,
\label{F2inv}\\
\tilde B_{h}(k_n;\o )&\equiv &
U^2  \l ^{n(1-h)-(h+{\cal Z}(h)+1) }
\tilde \C.B_{h}\Big( \frac{\o}{\o_{n;h}}\Big)\,, \label{B-inv}\quad  \\
\tilde T_{A;h}^{(\ell ) }(k_n;\o _a,\o _b )&\equiv & {i\over U} \l ^{n[h+{\cal
    Z}(h)]-\ell [2h+\C.Z(h)]}   \tilde \C.T_{A;h}^{(\ell ) }
\Big( \frac{\o _a}{\o_{n;h}},\frac{\o_b}{\o_{n;h}}\Big)\,,\qquad
\ell=-1\,,0\,,1\,,   \label{TA-inv}               \\
\tilde D_{A;h}^{(\ell ) }(k_n;\o _a,\o _b )&\equiv & ik_0\l ^{n[1+{\cal
    Z}(h)]-\ell [2h+\C.Z(h)]}   \tilde \C.D_{A;h}^{(\ell ) }
\Big( \frac{\o _a}{\o_{n;h}},\frac{\o_b}{\o_{n;h}}\Big)\,,
\label{D-inv}\\
\tilde E_{A;h}^{(\ell ) }(k_n;\o _a,\o _b )&\equiv & ik_0\l ^{n[1+{\cal
    Z}(h)]-\ell [2h+\C.Z(h)]}   \tilde \C.E_{A;h}^{(\ell ) }
\Big( \frac{\o _a}{\o_{n;h}},\frac{\o_b}{\o_{n;h}}\Big)\,,
\label{E-inv}
\EEA 
Here the characteristic frequency of $n$-shell on an $h$-slice is 
\BE \o_{n;h}\equiv U k_0\mu ^n\,,\qquad  \mu\equiv \l ^{1-h} \,,
\EE
$\l $ is a spacing parameter in the shell models determined as the
ratio between two consecutive shell momenta: $\l =k_{n+1}/k_{n}$. Recall 
that the dependence on a single $k_n$ is represents in the case of
3rd order functions a dependence on three consecutive shell momenta
$k_{n-1}$, $k_n$ and $k_{n+1}$.  The superscript $^{(\ell)}$ determines
the $k$-argument of the ``special leg'' accordingly to the convention of
Sect.~\ref{ss:def-triplex}.

In the dimensionless form Eqs.~\Ref.omega-s ~ are scale invariant,
independent of the shell number. In these Eqs.:
\BEA \label{B}
&& \C.B_h(\O )= {a\over  R_{_\C.Z}}V_{a;h}(\O)+b\,V_{b;h}(\O)+
cR_{_\C.Z}\,V_{c;h}(\O)\,,\quad \mbox{where}\quad 
R_{_\C.Z}\equiv \l ^{2+\C.Z(h)}\,, \\
\label{Va}
&& V_{a;h}(\O)=\int \frac{d \O _1 d \O_2 \d (\O+\O_1-\O_2)}{2\pi}
\tilde \C.F_{2;h}\Big (\frac{ \O_1}{\mu }\Big)
\tilde \C.F_{2;h}\Big (\frac{ \O_2}{\mu^2 }\Big)
 \tilde  \C.T_{A;h}^{(-1)}\Big (\frac{ \O_1}{\mu },\frac{ \O_2}{\mu
   }\Big)\,,  \\      \label{Vb}  
&& V_{b;h}(\O )=\int \frac {d \O _1 d \O_2 \d (\O+\O_1-\O_2)}{2\pi}
\tilde  \C.F_{2;h} ^* ( \O_1 \mu )
\tilde \C.F_{2;h}\Big (\frac{ \O_2}{\mu  }\Big)
 \tilde  \C.T_{A;h}^{(0)*}(\O_1,\O_2 )\,, 
\\    \label{Vc}  
&& V_{c;h}(\O)=\int \frac{d \O _1 d \O_2 \d (\O-\O_1-\O_2)}{2\pi}
\tilde \C.F_{2;h}(\O_1\mu )
\tilde \C.F_{2;h}(\O_2\mu^2 )
 \tilde  \C.T_{A;h}^{(+1)}( \O_2\mu , \O_1\mu )\ .
\EEA
The RHS of  Eq.~\Ref.eqTz  ~for $\C.T_{A;h}^{(0)}$ has the form:
\BEA\label{ED}
&& \tilde \C.E_h^{(0)}(\Om,\Op )
+ \tilde \C.D_h^{(0)}(\Om,\Op )={a\over  R_{_\C.Z}}W_{a;h}(\Om,\Op )
+cR_{_\C.Z}\,W_{c;h}(\Om,\Op ) \,, \qquad  \mbox{where}\\     \label{Wa} 
&& W_{a;h}(\Om,\Op ) = \int\frac{d\O}{2\pi}\Big\{\mu^3\tilde
\C.F_{2;h}(\O)  \tilde  \C.T_{A;h}^{(+1)*}\Big (\frac{ \O}{\mu },\frac{
  \Op}{\mu  }\Big)  
\tilde  \C.T_{A;h}^{(+1)}(\Om,\O)\\
& + &\mu^2\tilde   \C.F_{2;h}\Big({\O\over \mu} \Big)   
\tilde  \C.T_{A;h}^{(+1)*}\Big (\frac{ \O-\Om}{\mu },\frac{
  \Op}{\mu  }\Big) 
\tilde  \C.T_{A;h}^{(0)}(\Om,\O)         
+\mu \tilde                    
\C.F_{2;h}\Big({\O\over \mu^2}\Big)   
\tilde  \C.T_{A;h}^{(-1)*}\Big (\frac{ \Op}{\mu },\frac{
  \O}{\mu  }\Big)        
              \tilde  \C.T_{A;h}^{(0)}(\Om  ,\O-\Op )\Big\}\,,\nl
&& W_{c;h}(\Om,\Op ) = \int\frac{d\O}{2\pi}\Big\{\frac{1}{\mu^3}\tilde
\C.F_{2;h}(\O)  \tilde  \C.T_{A;h}^{(-1) }(\O \mu ,
  \Op\mu )  
\tilde  \C.T_{A;h}^{(-1)}(\O ,\Op)      \label{Wc}\\ 
& + &\frac{1}{\mu^2}\tilde   \C.F_{2;h}(\O \mu )   
\tilde  \C.T_{A;h}^{(-1)}[\Om ,(\Op-\O)\mu ]
\tilde  \C.T_{A;h}^{(0)}(\O,\Op)    
+\frac{1}{\mu} \tilde                    
\C.F_{2;h}(\O \mu^2  )   
\tilde  \C.T_{A;h}^{(+1)}(\O \mu ,
  \Om\mu  )        
              \tilde  \C.T_{A;h}^{(-1)}(\O+\Om ,\Op)\Big\}\ .\nonumber
\EEA  
%%%%%%%%%%%%%%%%%%%%%%%%%%%%%%%*********
\begin{multicols}{2}

\end{multicols}

\end{document}